\documentclass[aoas]{imsart}

\RequirePackage{amsthm,amsmath,amsfonts,amssymb}
\RequirePackage[authoryear]{natbib}
\RequirePackage[colorlinks,citecolor=blue,urlcolor=blue]{hyperref}
\RequirePackage{graphicx}
\usepackage{graphicx, verbatim}
\usepackage{booktabs,dcolumn,caption}
\usepackage{caption}
\usepackage{hhline}
\usepackage{graphicx, verbatim}
\usepackage{algorithm}
\usepackage{algpseudocode}
\usepackage{hyperref}
\usepackage{makecell}
\usepackage{tikz}
\usepackage{url}
\usepackage{bbm}
\usepackage{cases}
\usepackage{tikz}
\usepackage{verbatim}
\usepackage{float}
\usepackage{bbm}
\usepackage{csquotes}
\usepackage{multirow}

\begin{document}

\begin{frontmatter}
\title{Bayesian Data Synthesis and the Utility-Risk Trade-Off for Mixed Epidemiological Data}
\runtitle{Bayesian Data Synthesis}

\begin{aug}
\author[A]{\fnms{Joseph} \snm{ Feldman}\ead[label=e1]{joseph.r.feldman@rice.edu}}
\and 
\author[B]{\fnms{Daniel R.} \snm{Kowal}\ead[label=e2]{daniel.kowal@rice.edu}}
\address[A]{Department of Statistics,
Rice University,
\printead{e1}}

\address[B]{Department of Statistics,
Rice University,
\printead{e2}}

\end{aug}

\begin{abstract}
Much of the micro data used for epidemiological studies contain sensitive measurements on real individuals. As a result, such micro data cannot be published out of privacy concerns, and without public access to these data, any statistical analyses originally published on them are nearly impossible to reproduce. To promote the dissemination of key datasets for analysis without jeopardizing the privacy of individuals, we introduce a cohesive Bayesian framework for the generation of fully synthetic high dimensional micro datasets of mixed categorical, binary, count, and continuous variables. This process centers around a joint Bayesian model that is simultaneously compatible with all of these data types, enabling the creation of mixed synthetic datasets through posterior predictive sampling. Furthermore, a focal point of epidemiological data analysis is the study of conditional relationships between various exposures and key outcome variables through regression analysis. We design a modified data synthesis strategy to target and preserve these conditional relationships, including both nonlinearities and interactions. The proposed techniques are deployed  to create a synthetic version of a confidential dataset containing dozens of health, cognitive, and social measurements on nearly 20,000 North Carolina children.  
\end{abstract}

\begin{keyword}
\kwd{Copula}
\kwd{Data privacy}
\kwd{Factor model}
\kwd{Non-parametric regression}
\end{keyword}

\end{frontmatter}

\section{Introduction} \label{intro}

The combined impact of social and environmental stressors on childhood cognitive development remains understudied, limited in part by accessibility to rich, individual-level data. Of particular interest is the modeling of proxies for cognitive development, such as academic achievement, as potentially complex, nonlinear functions of different social and environmental exposures. In such work, the primary statistical tool used to study these relationships is regression analysis, which requires ample \emph{micro data} collected on individual subjects within a population \citep{miranda2007relationship,kowal2021fast}.  Such micro data often contains sensitive information that precludes its public release. While the inferences drawn from analyzing such micro data may have profound policy implications, the micro data itself often remain under lock and key out of concern for subject confidentiality, which undermines the scientific reproducibility of these analyses.

For this work, we examine a dataset built by linking three administrative datasets for the state of North Carolina (NC): {\bf detailed birth records}, which includes maternal demographics, maternal and infant health, and maternal obstetrics history for all documented live births in NC; {\bf blood lead surveillance} data from the state registry maintained by the Childhood Lead Poisoning Prevention Program of the Children’s Environmental Health Unit, Department of Health and Human Services in Raleigh, NC, which includes integer-valued blood lead levels; and {\bf end-of-grade (EOG) standardized testing data} from the NC Education Research Data Center of Duke University in Durham, which includes EOG reading and mathematics test scores from the 1995-1996 school year to the present, student identifying information, and data on demographics and socioeconomics.  Using residential addresses, these datasets are further linked with extensive exposure data, including environmental exposures (ambient air quality and temperature) and social exposures (indices for racial isolation and neighborhood deprivation); additional details are provided in the supplementary material as well as \cite{BVS, supp}. 

These micro data are high dimensional and consist of child-level measurements of various data types (categorical, binary, count, continuous). See Table~\ref{datad}. Several of these measurements are also highly sensitive, such as those taken on children's health, academic achievement, and socioeconomic status. While there are no unique identifiers in the data, other features in the dataset that contain high resolution geographic information connected to places of residence could be used by an ill-intended adversary to link individuals to these protected attributes with high accuracy. As a result, publishing these data could spawn a myriad of privacy and legal issues.

\begin{table}[H]
\centering
\caption{\label{datad} North Carolina dataset description}
\begin{tabular}{|p{1.35cm}|p{3.45cm}|p{4.75cm}|p{3.0cm}|}
\hline
\textbf{Type} & \textbf{Variable} & \textbf{Range/Values} & \textbf{Label}\\
\hline
\multirow{4}{*}{Categorical} & Mother's Race & White, Black,  Hispanic, Asian/Pacific Islander, Other & \verb|M_RACEGROUP| \\\cline{2-4}
    & Mother's Education &  No H.S. Diploma, H.S. Diploma, Some College/Associates Degree, Bachelor's or Higher & \verb|M_EDUCGROUP| \\
 \hline
\multirow{7}{*}{Binary} &Gender & Male, Female & \verb|MALE| \\\cline{2-4}
 & Prenatal Care & Yes, No& \verb|NOPNC| \\\cline{2-4}
 & Marital Status &Married, Not Married & \verb|NOTMARRIED|\\\cline{2-4}
& Smoker & Yes, No & \verb|SMOKER|\\ \cline{2-4}
& Econ. Disadvantaged  & Yes, No & \verb|ED|\\ \cline{2-4}
& Medicaid & Yes, No & \verb|MEDICAID|\\\hline 
\multirow{7}{*}{Integer} & EOG Reading Score & Integer 316-370 & \verb|ReadScal1| \\\cline{2-4}
& EOG Math Score & Integer 321-373 & \verb|MathScal1| \\\cline{2-4}
& Mother's Age (years) & Integer 15-40 & \verb|MAGE| \\\cline{2-4}
& Birth Weight Percentile & Integer 0-100 & \verb|BWTpctl_clin|\\\cline{2-4}
& Gestational Period (Weeks)& Integer 32-42  & \verb|GEST|\\\cline{2-4}
& Blood Lead Test Result & Integer 1-10  & \verb|PBresult|\\\hline
\multirow{12}{*}{Continuous}
& PM 2.5 (by trimester)  & PPM 5.60-31.06 & \verb|Ti_pm25_24hr| \verb|i = 1,2,3|\\ \cline{2-4}
& Acute PM 2.5 Exposure & PPM 6.026 - 17.891& \verb|PM25_30days_June|\\ \cline{2-4} 
& Chronic PM 2.5 Exposure & PPM 7.549 - 11.709 & \verb|PM25_1yr_June| \\ \cline{2-4}
& Racial Isolation at Birth & 0 - 1 & \verb|RI_nhb_Birth| \\ \cline{2-4}
& Racial Isolation at Test  & 0 - 1& \verb|RI_nhb_Educ| \\ \cline{2-4}
& Neighborhood Deprivation Index at Birth & -4.5174 - 11.3888 & \verb|NDI_Birth| \\ \cline{2-4} 
& Neighborhood Deprivation Index at Test & -4.17036 - 10.3524 & \verb|NDI_Educ|\\ \hline
\end{tabular}
\end{table} 

To address this issue, we develop a Bayesian framework for the construction of fully synthetic datasets that is compatible with mixed data types and able to scale to high dimensions. It is important to emphasize the ability of this framework to seamlessly incorporate modeling of \emph{unordered categorical variables} alongside other data types, as unordered categorical variables, especially race or ethnicity, are critical in health disparities research and practice.

We apply our method to create a fully synthetic version of the aforementioned North Carolina dataset.  In addition to providing consistent univariate, bivariate, and multivariate relationships in the synthetic data, our method ensures that outcomes measuring cognitive development of subjects display similar relationships with key exposure variables observed in the original data. Given the aims of studying the original dataset, maintaining these structures is the most desired feature of any synthetic version. Because the synthetic data comprises no real individuals, it does not jeopardize the privacy of individuals in the original dataset. In addition, we confirm minimal attribute disclosure risks via several tests. As such, researchers may request access to the synthetic dataset at \url{https://doi.org/10.25614/synthetic_data}, and unrestricted public access is currently pending approval.

The literature on the construction of fully synthetic datasets to preserve confidentiality dates back at least to \cite{rubin1993statistical} who approached data synthesis as a problem of multiple data imputation.  Under this \textit{fully synthetic} framework, the data disseminating agency views synthetic observations as missing observations from a population. A generative model is fit to the confidential data, and synthetic observations are randomly sampled values from multiple imputations generated from a predictive distribution.

Our approach falls more in line with \textit{partial synthesis} \citep{little1993statistical} - though we synthesize all variables in the dataset - which creates synthetic observations by directly sampling from a generative model conditioned on confidential data \citep{drechsler2018some}. For this work, our method and its competitors are technically partial synthesizers, though we refer to each in their ability to create completely synthetic datasets, since all variables in the confidential set are synthesized.  See \cite{raab2016practical} for further clarification on inference for partially versus fully synthetic data.

A popular approach for generation of completely synthetic datasets is known as  sequential  conditioning: using the relationship $P(Y_{1},\dots,Y_{p}) = P(Y_{1}) \prod_{j = 2}^p P(Y_{j}\mid Y_{1},\dots, Y_{j - 1})$, separate models are fit for each $P(Y_{j}\mid Y_{1},\dots, Y_{j - 1})$  using confidential data \citep{kinney2011towards}. Then, a synthetic version, $Y_{j_{syn}}$, is realized through bootstrapping, posterior predictive sampling, or density estimation. \cite{reiter2005releasing} used parametric linear regression models at each iteration of the conditioning. More recently, nonparametric, tree-based methods have emerged  \citep{reiter2005using,caiola2010random}, which can provide a more effective approach for modeling nonlinear relationships in the data.  

The primary limitation of the  sequential conditioning approach is the need to choose the variable ordering for data synthesis. As demonstrated in Section~\ref{results}, the utility and privacy of the synthetic data are highly sensitive to the ordering. Ideally, one might consider optimizing the ordering for maximal utility and privacy. Yet this becomes intractable for moderate to large $p$, since the number of orderings to consider is $p!$. In our setting, the dimension is $p=23$, which is far too large for a combinatorial search.

In our case, the dataset to be synthesized features categorical (both ordered and not), binary, count, and continuous measurements on North Carolina children.  From a joint modeling perspective, this necessitates a generative model simultaneously compatible with each type.  Existing Bayesian generative models are able to estimate joint distributions of decidedly non-Gaussian data types, potentially on mixed scales. For instance,  \cite{murray2016multiple} develop a hierarchical mixture model specifically aimed at multiple imputation for mixed continuous and categorical outcomes. \cite{dunson2009nonparametric} provides a joint model for multivariate categorical data, and \cite{hu2014disclosure} uses this model for the generation of synthetic multivariate categorical data.  \cite{quick2015bayesian} use Bayesian marked point process modeling to create synthetic datasets for epidemiological analyses that preserve spatial dependence structures. Closely aligned with this work, \cite{hoff2007extending} and \cite{murray2013bayesian} estimate a semiparametric Gaussian copula for data of mixed continuous, count, and ordinal types through a marginal likelihood called the extended rank likelihood.  However, none of these methods simultaneously deal with categorical, binary, count, and
continuous variables as in our NC dataset.

Our main contribution is in the development of a Bayesian semiparametric generative model for the joint distribution  of categorical, binary, count, and continuous random variables. The model is based on a semiparametric Gaussian copula, which provides a convenient platform for linking univariate marginal distributions under a multivariate dependence structure to induce a joint distribution.  A synthetic dataset can then be constructed by simulating samples from the posterior predictive distribution. Under this framework, variables are synthesized jointly, which obviates the need to declare an ordering for data synthesis.  Additionally, our data synthesizer is capable of coherently and accurately capturing the relationships among categorical, binary, count, and continuous variables. Therefore, a notable achievement of this work is in the development of a joint model for unordered categorical, binary, count, and continuous variables, a previously significant hurdle for existing Bayesian methods.

The North Carolina dataset is particularly useful for deriving insights into the relationships between social and environmental exposures and EOG test scores. Consequently, we design the synthetic data generating process to target and capture the regression associations for modeling EOG test scores using the other demographic, socioeconomic, and exposure variables.  To do so, we incorporate a nonparametric and nonlinear Bayesian regression model using the EOG test scores as the response variable. This approach capitalizes on the benefits of  sequential conditioning  by directly identifying the key response variables of interest, yet avoids the difficulties in selecting an ordering by generating the remaining variables---all but two (EOG reading and mathematics scores)---jointly using our semiparametric copula model.

Ultimately, we are able to demonstrate several key properties of our synthetic data. The semiparametric and multivariate components provide remarkable consistency for marginal distributions and in crucial bivariate relationships. Perhaps more important to our application, we demonstrate the high utility datasets produced under our method when summarizing the inference from linear regression models for both synthetic and confidential data. Such validation is vital if the synthetic data are to be used by additional investigators beyond the primary managers of confidential databases. We must ensure that the synthetic data "tell the same story" as the confidential data.

 Finally, we introduce a general methodology for examining the utility-risk trade-off for synthetic data under the common practice of generating and releasing $m$ datasets. We pose the privacy risk as a function of several decisions that the data disseminating agency must make when releasing synthetic data publicly, and evaluate these risks in tandem with utility for each of the synthesizers considered. We show that under varying conditions, the proposed approach yields synthetic data that is similarly protective and more useful than its competitors.   

The organization of this paper is as follows. In Section~\ref{sgc}, we introduce a semiparametric Gaussian copula model for estimating a joint distribution. Section~\ref{erpl} describes our main methodological innovation: the \textit{extended rank-probit likelihood} as a model for unordered categorical, binary, count, and continuous variables. Section~\ref{sec4} includes the modeling details, the MCMC sampling algorithm, and the data synthesis. Section~\ref{nonlinear} details the modifications for nonlinear regression. The results and the utility-risk trade-off are presented in Section~\ref{results}. We conclude in Section~\ref{conclude}.

\section{A Semiparametric Gaussian Copula}\label{sgc}
To generate a fully synthetic dataset, we first build a model for the joint distribution of the confidential data.  Copula models provide an effective strategy: by linking marginal distributions with a multivariate dependence structure, copula models can preserve both marginal properties and multivariate relationships.  By Sklar's Theorem \citep{sklar1959fonctions}, the joint distribution of a $p$-dimensional random vector $Y = (y_{1}, \dots, y_{p})$ can be expressed through the univariate marginals $F_{j}, j = 1,\dots, p$ and a copula $\mathbb{C}$:
\[F(y_{1},\dots, y_{p}) = \mathbb{C}\{F_{1}(y_{1}),\dots, F_{p}(y_{p})\}. \]
For computational convenience and modeling flexibility, we build upon the Gaussian copula
\begin{equation}\label{eq1}
    \mathbb{C}(u_{1},\dots, u_{p})= \Phi_{p}\{\Phi^{-1}(u_{1}), \dots \Phi^{-1}(u_{p})\}
\end{equation}
where $u_j \in [0,1]$ for $j=1,\ldots,p$ and $\Phi_{p}$ is the cumulative distribution function (CDF) of a $p$-dimensional Gaussian random vector with correlation matrix $\boldsymbol{C}$. The joint distribution of $Y$ is derived by combining \eqref{eq1} with univariate marginal distributions $\{F_{j}\}_{j=1}^p$. Since $F_{j}(y_{j}) \sim \mbox{Uniform}(0,1)$, the joint distribution is
\begin{equation} \label{2}
    F(y_{1}, \dots, y_{p}) =\Phi_{p}[\Phi^{-1}\{F_{1}(y_{1})\}, \dots, \Phi^{-1}\{F_{p}(y_{p})\}]
\end{equation}
which provides a generative model for multivariate data $Y$. 

A Bayesian approach requires a probability model for marginal distributions $\{F_j\}_{j=1}^p$ and the parameters that govern the Gaussian copula---namely, the correlation matrix $\boldsymbol{C}$. Given samples from the posterior distribution of these parameters, it is then possible to simulate posterior predictive samples, enabling construction of a fully dataset. In particular, the data generating process is defined by (i) sampling latent data $\boldsymbol z \sim N_{p}(\boldsymbol 0, \boldsymbol{C})$ and (ii) computing the marginals $y_{j} = F_{j}^{-1}\{\Phi(z_{j})\}$ for $j=1,\ldots,p$. The latent Gaussian random variables $\boldsymbol z$ capture the dependence structure among the variables, while the marginal inverse CDFs $\{F^{-1}\}_{j=1}^{p}$  link these dependent random variables to the correct scale of confidential data. Since the marginal distributions $F_j$ can be estimated accurately using empirical CDFs or other marginal models even for small to moderate sample sizes, the important modeling task centers on the dependence structure $\boldsymbol{C}$. 

 Given data $\{y_{ij}\}$, a natural semiparametric strategy for inference on  $\boldsymbol{C}$ is to compute psuedo-data $z_{ij} = \Phi^{-1}\{\hat{F_{j}}(y_{ij})\}$ for observations $i=1,\ldots,n$, where $\hat{F}_{j}$ is an estimate of each marginal CDF, and to perform inference on the model $\boldsymbol z_{i} \sim N_{p}(\boldsymbol 0, \boldsymbol{C}) $ independently for $i=1,\ldots,n$. However, problems arise for data of mixed types. Consider maximum likelihood estimation of $\boldsymbol{C}$: when the marginal distributions $F_j$ are continuous, the estimator is consistent; yet when the marginal distributions are discrete, the psuedo-data transformation changes only the sample space and not the data distribution so the resulting estimator is inconsistent \citep{hoff2007extending}. These issues persist for Bayesian models.

For continuous, count, and ordinal variables, \cite{hoff2007extending} proposes a remedy based on the rank likelihood. Since CDFs are non-decreasing, observing $y_{ij} < y_{lj}$ implies that $ z_{ij} < z_{lj}$. We can make this partial ordering precise for each variable $j$:
\begin{equation}\label{3a}
   \boldsymbol{D(y_{j})} =  \{\boldsymbol{z}\in \mathbb{R}^{n}: y_{ij}< y_{lj} \implies z_{ij} < z_{lj}, \forall i\neq l \in 1, \dots, n\},
\end{equation}
so $\boldsymbol{D(y_{j})}$ is the set of all values of $\boldsymbol{z_{j}}$ that match the ordering of the confidential data $\boldsymbol{y_{j}}$.  For $n$ observations $\boldsymbol{Y} = (Y_1, \dots, Y_{n})'$, the latent variables $\boldsymbol{Z} = (Z_1, \dots, Z_{n})'$ must satisfy the event
$
    \boldsymbol{D} = \{ \boldsymbol{Z} \in \mathbb{R}^{n \times p}: max\{z_{kj}: y_{kj}< y_{ij}\} <z_{ij}< min\{z_{kj}: y_{ij}< y_{kj}\}, \forall j = 1,\dots,p\}. 
$ 
The full data likelihood can thus be decomposed
\begin{align}
    P(\boldsymbol{Y} \mid \boldsymbol{C}, F_{1},\dots,F_{p}) &= P(\boldsymbol{Y}, \boldsymbol{Z} \in \boldsymbol{D} \mid \boldsymbol{C}, F_{1}, \dots, F_{p}) \label{eq7}\\ 
    &= P(\boldsymbol{Z} \in \boldsymbol{D} \mid \boldsymbol{C}) \times P( \boldsymbol{Y} \mid  \boldsymbol{Z} \in \boldsymbol{D}, \boldsymbol{C}, F_{1}, \dots, F_{p})\label{eq8.1}
\end{align}
The equivalence in \eqref{eq7} is true by construction, since observing $\boldsymbol{Y}$ implies that $\boldsymbol{Z} \in \boldsymbol{D}$, and \eqref{eq8.1} follows because the event $\boldsymbol{Z} \in \boldsymbol{D}$ is independent of the marginal distributions $F_{1},\dots, F_{p}$. \cite{hoff2007extending} proposes to estimate the copula parameters $\boldsymbol{C}$ by treating  $P(\boldsymbol{Z} \in \boldsymbol{D} \mid \boldsymbol{C})$ as the likelihood, which he refers to as the \emph{extended rank likelihood}. With this likelihood, \cite{hoff2007extending} models $\boldsymbol{C}$ using an inverse-Wishart prior;  \cite{murray2013bayesian} adopt a similar approach based on a factor model. Although the extended rank likelihood is a rank-based approximation for the true likelihood, it nonetheless contains much of the information about $\boldsymbol{C}$. \cite{murray2013bayesian} confirm this intuition by showing strong posterior consistency for $\boldsymbol{C}$ in a reduced rank setting.

\section{The Extended Rank-Probit Likelihood}\label{erpl}
 As a preliminary, we distinguish between ordered and unordered categorical variables. Both variable types are qualitative descriptors, but ordered categorical variables possess an implicit ranking, while unordered categorical variables do not.  With this distinction in mind, there are three crucial limitations in the Gaussian copula model and the extended rank likelihood. These limitations are described here and resolved in subsequent sections

First, an incongruity arises in applying the extended rank likelihood for data that contain unordered categorical variables. The extended rank likelihood is built upon (i) inverse marginal CDFs and (ii) the ordering of the confidential data. Unordered categorical variables have neither a well-defined inverse CDF nor a natural ordering. The ability to model unordered categorical variables is crucial. For example, race or ethnicity is a critical factor in health and health disparities research and practice. Use of the extended rank likelihood here would require an ordering of the races---a task that is both unethical and nonsensical.

A simple workaround is to one-hot-encode each categorical variable with $k$ levels as $k-1$ binary variables, and then incorporate these variables into the extended rank likelihood. This solution is unsatisfactory. By treating a categorical variable as $k-1$ separate binary variables, the data generating process does not necessarily respect the fundamental constraint that each individual belongs to exactly one category. As a result, a synthetic dataset is likely to violate this constraint, possibly many times. For our dataset, this workaround generates nearly 500 synthetic individuals that belong to more than one race category. More subtly, the omission of this constraint has modeling implications for the correlation structure. Inference on $\boldsymbol{C}$ will be restricted to $k-1$ levels of any given categorical variable, which requires selection of a ``base" category. Resulting inference, including data synthesis, will depend on the choice of base category for each categorical variable. 


 Second, \cite{hoff2007extending} acknowledges that the extended rank likelihood performs poorly for ordinal variables with few levels. In practice, we find that these variables are better modeled as \emph{unordered} categorical variables using our proposed approach, which is confirmed for both simulated data and the North Carolina dataset (see Section~\ref{results}).

Third, the correlation matrix $\boldsymbol{C}$  only captures linear dependencies on the latent scale. As a result, the Gaussian copula may be an overly simplistic model for outcomes of interest for which nonlinear dependencies need be preserved. We address this challenge in Section~\ref{nonlinear}. 

In what follows, we propose an \textit{extended rank-probit likelihood}, which more naturally models categorical and ordinal data with few levels. We use this likelihood to define a new Gaussian copula model for mixed categorical, binary, ordinal, count, and continuous data.

\subsection{A Marginal Likelihood for Unordered Categorical Variables}
 Suppose that a categorical variable in the dataset, $\boldsymbol{y}_{j}$, posseses $k$ distinct levels. We represent this variable as $k$ binary variables $\gamma_{j_{l}} \in \{0,1\}, l = 1,
\dots, k$, such that \[y_{ij} = m \iff \{\gamma_{ij_{m}} = 1\} \cap \{\gamma_{ij_{l}} = 0\}, \forall l \neq m.\] 
These binary variables can be expressed using a latent Gaussian representation akin to probit regression \citep{albert1993bayesian}. Specifically, we introduce latent variables $(z_{ij_{1}},\dots, z_{ij_{k}})$ such that the event $\{\gamma_{ij_{m}} = 1\} \cap \{\gamma_{ij_{l}} = 0\}, \forall l \neq m$ is equivalent to 
\begin{equation}\label{DO-Probit}
    (z_{ij_{1}},\dots, z_{ij_{k}}) \in \boldsymbol{d'(y_{j})} \equiv \cup_{m=1}^{k}\{\boldsymbol{z_{i}} \in \mathbb{R}^{k}: z_{im} >0, z_{il}<0, l \neq m\}
\end{equation}
analogous to $\boldsymbol{D(y_{j})}$ in \eqref{3a}. The latent variables in \eqref{DO-Probit} describe the inclination of observations to belong to one level over another. With this representation, the multinomial data do imply a partial ordering among latent variables, but the ordering comes \textit{within} each observation: the ranking occurs among the levels of a categorical variable, and this ranking is unique to the individual. This is contrary to what is induced by the rank likelihood, where a ranking is induced \textit{among} individuals through a comparison of numerical quantities observed in the population. We can then combine this latent variable representation of categorical data with the rank-based representation of numerical measurements, which we refer to as the   \textit{extended rank-probit likelihood}.

Consider a dataset with $p$ columns comprised of $q$ categorical variables, each with a potentially unique number of levels $k_{1},\dots, k_{q}$, and $r$ variables that are ordinal, count, continuous, or binary. For each categorical variable, we apply the latent binary representation from \eqref{DO-Probit} and aggregate across the entire dataset: 
$\boldsymbol{D'}_{c} = \{\boldsymbol{Z}^{n \times k_{c}}: \cup_{j =1}^{k_{c}}z_{ij}> 0, z_{il}<0, l \neq j\}$ for $c = 1,\dots, q$. 
Across all $q$ categorical variables, the observed group memberships in the dataset satisfy the event 
$ \boldsymbol{D'} = \cup_{c = 1}^{q}\boldsymbol{D'}_{c}$. 
As was the case with the rank likelihood, the probability that $\boldsymbol{Z}_{q}$ satisfies $\boldsymbol{D'(\boldsymbol{Y_{q}})}$ does not depend on the marginal distributions of the categorical variables, which allows us to join this event with the event $ \boldsymbol{D}$. More formally, we have that observing the full dataset $\boldsymbol{Y} \in \mathbb{R}^{n \times p^{*}}$, where $p^{*} = r + \sum_{c = 1}^{q}k_{c}$, implies that $\boldsymbol{Z}$ must satisfy the new event $\boldsymbol{E} = \boldsymbol{D} \cup \boldsymbol{D'}$. Like \eqref{eq7}-\eqref{eq8.1}, we can decompose the full data likelihood using this new event:
    \begin{align}
    P(\boldsymbol{Y} \mid \boldsymbol{C}, F_{1},\dots,F_{p}) &= P(\boldsymbol{Y}, \boldsymbol{Z} \in \boldsymbol{E} \mid \boldsymbol{C}, F_{1}, \dots, F_{p}) \label{eq11}\\ \label{eq17} &= P(\boldsymbol{Z} \in \boldsymbol{E} \mid \boldsymbol{C}) \times P( \boldsymbol{Y} \mid  \boldsymbol{Z} \in \boldsymbol{E}, \boldsymbol{C}, F_{1}, \dots, F_{p}) 
    \end{align}
where the equivalence in \eqref{eq11} once again arises since observing $\boldsymbol{Y}$ implies that $\boldsymbol{Z}$ satisfies the event $\boldsymbol{E}$. In the Gaussian copula, the parameter of interest is $\boldsymbol{C}$, and in the decomposition of the full data likelihood in \eqref{eq17}, we see that the left term depends solely on $\boldsymbol{C}$. As in \cite{hoff2007extending}, we proceed to estimate the Guassian copula  parameters solely based on the marginal likelihood, $P(\boldsymbol{Z} \in \boldsymbol{E} \mid \boldsymbol{C})$, which is the \textit{extended rank-probit likelihood}.
\subsection{Adjusting the Gaussian Copula Sampling Model}

 The event $\boldsymbol{D_{c}'(y_{c})}$ is identical to the set restriction imposed for the data augmentation in the diagonal orthant multinomial probit model of \cite{johndrow2013diagonal}. In this model, it is possible to derive a convenient link function whereby the mean of the multivariate normal distribution governing latent $\boldsymbol{z}$ gives class probabilities for each observation. Specifically, the event $\boldsymbol{D_{c}'(y_{c})}$ induces the following probability distribution on $\boldsymbol{y_{c}}$:
\begin{equation}
    P( \boldsymbol{y_{c}}  = h) = P(z_{h} >0, \{z_l<0\}_{l \ne h}). \label{eq18}
\end{equation}
This general format lends insight into how we can interpret latent variables corresponding to categorical levels that satisfy the extended rank-probit likelihood.  For categorical data, there is no relationship between latent variables and realizations of observed data through a marginal CDF. Instead, each categorical observation is represented through a $k_{c}$-dimensional latent vector which encodes the class membership probabilities for that observation. 

This results in a simple modification to the Gaussian copula model, namely that we must estimate an intercept, $\boldsymbol{\alpha}$, for the $p^{*}$-dimensional Gaussian distribution that characterizes the multivariate dependence structure in our data. This intercept is non-zero only for components of the latent vector corresponding to levels of a categorical variable. The reason for this is simple: \eqref{eq18} tells us that if we were to leave the mean vector to be \textbf{0}, our model would imply equal probability among the levels of each categorical variable.

\section{Bayesian Estimation and Data Synthesis} \label{sec4}
\subsection{A Factor Model for the Extended Rank-Probit Likelihood}
Bayesian inference and synthetic data generation for the extended rank-probit likelihood requires a model for the latent Gaussian variables $\boldsymbol{z}$. Factor models offer a natural approach: they capture dependence among high dimensional data through a parsimonious low-rank model and computationally scalable posterior sampling  algorithms. 
Specifically, our model is given by:
\begin{equation} \label{eq22}
    \boldsymbol{z_{i}} = \boldsymbol{\alpha} + \boldsymbol{\Lambda\eta_{i}} + \boldsymbol{\epsilon_{i}}, \quad \boldsymbol{\epsilon_{i}} \stackrel{iid}{\sim} N(\boldsymbol{0}, \boldsymbol{\Sigma})
\end{equation}
where $\boldsymbol{\alpha}$ encodes categorical probabilities, $\boldsymbol{\Lambda}$ is a $p^{*} \times k$ matrix of factor loadings, $\boldsymbol{\eta_{i}}$ a $k \times 1$ vector of factors, and  $\boldsymbol{\Sigma} = diag(\sigma_{1}^{2}, \dots, \sigma_{p^{*}}^{2})$. 

The dependence among the latent variables is captured in the lower dimensional vector of factors, $\boldsymbol{\eta_{i}}$, usually with $k < < p^{*}$. By marginalizing over $\boldsymbol{\eta}$, the latent variables satisfy $\boldsymbol{z}_{i} \sim N(\boldsymbol{\alpha}, \boldsymbol{\Omega})$ with reduced rank covariance $\boldsymbol{\Omega} = \boldsymbol{\Lambda \Lambda ' + \Sigma}$. The advantage of modeling the data on the covariance scale, rather than using correlations, is computational simplicity. Priors for $\boldsymbol{\lambda}, \sigma_{j}, \text{and} \ \boldsymbol{\eta_{i}}$ enjoy  conjugacy under a Gaussian likelihood. As a result, a simple and effective Gibbs sampling algorithm may be developed. In contrast, correlation matrices have rigid structure; the diagonal terms must be one and the off diagonal terms must be between zero and one in absolute value. A comparable factor model for a correlation matrix would require alternative priors and more complex sampling algorithms. Still, posterior inference for the correlation matrix is available under \eqref{eq22}: given posterior samples of the covariance matrix $\boldsymbol \Omega$, we can easily rescale   $\boldsymbol{C}_{ij} = \omega_{ij} / \sqrt{\Omega_{ii}\Omega_{jj}}$ and similarly $\tilde{\alpha}_{j} = \alpha_{j} / \sqrt{\Omega_{jj}}$.

We specify the following priors for the non-zero components of $\boldsymbol{\alpha}$, $\sigma_{j}^{-2}$, and $\boldsymbol{\eta_{i}}$:
\begin{equation*}
 \alpha_{j} \sim N(0,1), \ \sigma_{j}^{-2} \sim IG(a_{\sigma}, b_{\sigma}), \ \boldsymbol{\eta_{i}} \sim N(\boldsymbol 0,\boldsymbol{I_{k}})
\end{equation*}
independently. For the elements of $\boldsymbol{\Lambda}$, we utilize the multiplicative gamma process prior of \cite{bhattacharya2011sparse}: 
$[\lambda_{j,h} \mid \phi_{jh}, \tau_{h}] \sim N(0,\phi_{jh}^{-1}\tau_{h}^{-1})$ independently with local scale parameters $\phi_{jh}\sim Gamma(\nu/2, \nu/2)$ and global scale parameters $\tau_{h} = \prod_{l =1}^{h}\delta_{l}$ for 
$\delta_{1} \sim Gamma(a_{1},1)$ and 
$\delta_{l} \sim Gamma(a_{2},1), \ l \geq 2$.  Because of the stochastically increasing nature of $\tau_{h}$ (under the constraint that $a_{2} >1$), this prior favors lower dimensional representations and specifically lower rank for the covariance $\boldsymbol \Omega$. The parsimony afforded by this prior on the factor loadings allows estimation to scale well with $p$, making joint modeling, and therefore synthesis, of high dimensional and highly correlated datasets more feasible. 

For the marginal CDFs $F_j$ (for non-categorical variables) in the extended rank-probit likelihood, we substitute an estimate using the empirical CDFs $\hat F_j$, as in \cite{hoff2007extending} and \cite{murray2013bayesian}. This choice ensures that data realized through the generating model comes on the correct scale with consistent marginal properties, and can be re-scaled to avoid infinities. For continuous variables, we estimate the CDFs using a kernel smoother to ensure unique values in the data synthesis as an additional layer of privacy protection. More broadly, the extended rank-probit likelihood is compatible with any marginal model for each (non-categorical) variable $j$.

\subsection{MCMC Sampling Algorithm}
The MCMC sampling algorithm proceeds as follows, and is decomposed into three blocks for clarity:
\begin{enumerate}
    \item \textbf{Sample the Factor Model Parameters}:
    \begin{itemize}
        \item $\lambda_{j,-} \mid - \sim N((\boldsymbol{D_{j}}^{-1} + \sigma_{j}^{-2}\boldsymbol{\eta}^{T}\boldsymbol{\eta})^{-1}\boldsymbol{\eta}^{T}\sigma_{j}^{-2}(\boldsymbol{z_{j}} - \alpha_{j}))$, 
        where $\boldsymbol{D_{j}}^{-1} = diag(\phi_{j1}\tau_{1}, \dots, \phi_{jk}\tau_{jk})$, $\boldsymbol{z_{j}} = (z_{1j}, \dots z_{nj})^{T}$, and $\boldsymbol{\eta} = (\eta_{1j}, \dots \eta_{nj})^{T}$, for $j = 1, \dots, p^{*}$
        \item $\sigma_{j}^{-2} \mid - \sim Gamma(a_{\sigma} + \frac{n}{2}, b_{\sigma} + \frac{1}{2}\sum_{i = 1}^{n}\{z_{ij} - (\alpha_{j} + \lambda_{j}^{T}\eta_{ij})\}^{2}$, for $j = 1, \dots, p^{*}$
        \item $\eta_{i}\mid - \sim N_{k}(\boldsymbol{\boldsymbol{I_{k}}} + (\boldsymbol{\Lambda^{T}\Sigma^{-1}\Lambda)^{-1}}\boldsymbol{\Lambda^{T}\Sigma^{-1}}(\boldsymbol{z_{i} - \alpha}), (\boldsymbol{I_{k}} + \boldsymbol{\Lambda \Sigma^{-1}\Lambda})^{-1})$, where $\boldsymbol{z_{i}} = (z_{i1}, \dots, z_{ip^{*}})$, for $i = 1, \dots, n$
        \item $\phi_{jh} \mid - \sim Gamma( \frac{\nu + 1}{2}, \frac{\nu + \tau_{h}\lambda_{jh}^{2}}{2})$, for $j = 1, \dots, p^{*}, \ h = 1, \dots, k$ \\
        \item $\delta_{1} \mid - \sim Gamma(a_{1} + \frac{p^{*}k}{2}, 1 + \frac{1}{2}\sum_{l = 1}^{k}\tau_{l}^{(1)}\sum_{j =1}^{p}\phi_{jl}\lambda_{jl}^{2})$, and for $h \geq 2$ \\
        $\delta_{h} \mid - \sim Gamma(a_{1} + \frac{p^{*}(k -h + 1)}{2}, 1 + \frac{1}{2}\sum_{l = 1}^{k}\tau_{l}^{(h)}\sum_{j =1}^{p}\phi_{jl}\lambda_{jl}^{2})$, where $\tau_{l}^{h} = \prod_{t = 1, t \neq h}^{l} \delta_{t}$, for $h = 1, \dots, k$
    \end{itemize}
    \item \textbf{Sample the intercepts} $\boldsymbol{\alpha_{j}}$ for columns corresponding to categorical variable levels:
    
    \begin{itemize}
        \item $\alpha_{j}\mid - \sim N((n\sigma_{j}^{-2} + 1)^{-1}\sigma_{j}^{-2}\sum_{i = 1}^{n}\sum_{h =1}^{k}(z_{ij} - \lambda_{jh}\eta_{ih}), (n\sigma_{j}^{-2} + 1)^{-1})$
    \end{itemize}
    
    \item \textbf{Sample $\boldsymbol Z$:} for each column, sample $z_{ij}$ from a truncated normal, with lower and upper bounds for each observation specified by the extended rank-probit likelihood:
    \begin{itemize}
        \item $z_{ij} \mid - \sim TN( \alpha_j + \sum_{h =1}^{k}\lambda_{jh}\eta_{ik},\sigma_{j}^{1},z_{ij}^{l}, z_{ij}^{u})$, where $TN(\mu, \sigma^{2},a,b)$ denotes a truncated univariate normal with  mean $ \mu$, variance $\sigma^{2}$, lower truncation $a$, and upper truncation $b$. 
For ordinal, count, and continuous variables, the truncation limits are $z_{ij}^{l} = max\{z_{-ij}: y_{-ij}<y_{ij}\}$, and $z_{ij}^{u} = min\{z_{-ij}: y_{-ij}>y_{ij}\}$, where $z_{-ij} = \boldsymbol{z_{j}}\setminus z_{ij}$. For columns corresponding to categorical levels, the upper and lower truncation limits are
    \begin{equation}\label{trunc}
        z_{ij}^{l} = \begin{cases} 0, & y_{ij} = 1\\ -\infty, & y_{ij} = 0 \end{cases}, \quad    \quad          
        z_{ij}^{u} = \begin{cases} \infty, & y_{ij} = 1\\ 0, & y_{ij} = 0\end{cases}
    \end{equation}
    \end{itemize}

\end{enumerate}
Most notably, the factor model parameters and the intercepts benefit from the conditional Gaussianity in \eqref{eq22}, while the latent data imputation of $\{z_{ij}\}$ consists of univariate truncated normal sampling steps. In conjunction, these components produce an MCMC algorithm that is surprisingly scalable given the complexity and dimensionality of the model. 

\subsection{Data Synthesis} \label{synthesis}
Posterior predictive sampling is used to construct a fully dataset. In particular, given draws from the posterior distribution of $z_{ij}$ via \eqref{eq22} using the aforementioned sampling algorithm, posterior predictive draws for the non-categorical variables are simply given by $ \tilde y_{ij} = \hat F_{j}^{-1}\{\Phi(z_{ij})\}$. As a result, posterior predictive samples can be generated rapidly for these variables. 

However, posterior predictive draws for the categorical variables present additional challenges due to the  sign constraints under the rank-probit likelihood.  For each synthetic observation $i$, the latent variables $z_{ij_1}, \ldots, z_{ij_k}$ for categorical variable $j$ must satisfy the restriction that one and only one of these components is positive, which corresponds to the assigned category level. Since the full latent vector $\boldsymbol z$ may contain multiple categorical variables as well as a variety of non-categorical variables, enforcing such a constraint in the presence of the multivariate dependence implied by \eqref{eq22} creates significant computational issues. Note that this issue persists only for posterior \emph{predictive} sampling: in the case of posterior sampling of $z_{ij_1}, \ldots, z_{ij_k}$, the observed data $\{y_{ij}\}$ inform exactly which of these components must be positive, while the remaining components are constrained to be negative. In that case, samples are generated from independent and univariate truncated normals. By comparison, for posterior predictive sampling, we only know that exactly one of these components---but not \emph{which} component---must be positive. Hence, the constraint region is more difficult to enforce. 

To circumvent these challenges, we design a posterior predictive sampling approach that generates joint realizations $\tilde y = (\tilde y_{cat}, \tilde y_{-cat})$ by (i) marginally simulating the categorical variables $\tilde y_{cat}$ and (ii) simulating the non-categorical variables $\tilde y_{-cat} | \tilde y_{cat}$ conditionally on the categorical variables. The main advantages follow from (ii). First, given the categorical realizations $\tilde y_{cat}$, the corresponding latent variables $\tilde z_{cat}$ are generated exactly as in the MCMC algorithm. More specifically, $\tilde y_{cat}$ determines precisely which of the components of $z_{cat}$ must be positive, while the others must be negative. Hence, simulating $z_{cat} | \tilde y_{cat}, -$ from the full conditional distribution only requires sampling from  a multivariate truncated normal distribution, a task made simple by the availability of several existent software packages such as \verb|TruncatedNormal|. Second, given these draws of the latent $z_{cat}$, the augmented full conditional distribution $z_{-cat} | z_{cat}, -$ simply modifies \eqref{eq22} using standard results for Gaussian conditionals. 

For the marginal sampling of $\tilde y_{cat}$ in (i), we adopt a fast and simple approach similar to the empirical CDF estimates of $F_j$ for non-categorical variables. Let $\boldsymbol{\hat{p}}$ denote the vector of empirical probabilities over the union of all categorical variables. For example, the North Carolina dataset featured two variables that we modeled as categorical, mother's race and education level, with four and five categories, respectively.  Consequently, $\boldsymbol{\hat{p}}$ is a 20-dimensional vector of empirical probabilities. 

Using this empirical estimate, we simply generate $\tilde y_{cat}$ from a multinomial distribution with probability vector $\boldsymbol{\hat{p}}$  using the \verb|sample| function in \verb|R|. This step  ensures that categorical observations belong to exactly one level and guarantees that we observe consistent marginal properties for these categorical variables.  In scenarios where the number of categories from the union of categorical variables is much higher, or if there are structural zeros in the resulting contingency table, we may substitute any marginal model to estimate $\boldsymbol{\hat{p}}$, enabling us to scale this approach to datasets containing many categorical variables.

Note that this modification does not preclude the inclusion of the categorical variables in the extended rank-probit likelihood and factor model \eqref{eq22}: the sampling step $\tilde y_{-cat} | \tilde y_{cat}$ relies on the learned associations from the latent $z_{-cat} | z_{cat}, -$ under the proposed model.  In addition, we mention here that we modeled mother's education level as unordered categorical because there are only four levels for this variable. This decision was motivated by the results from our simulation study in~\ref{sims}, in which we provide evidence that ordinal variables with few levels ($k < 10$) are better modeled as categorical under the RPL.

This approach maintains the core of a semiparametric Gaussian copula model: marginal distributions are estimated empirically and the joint distributions are modeled via latent Gaussian variables. The primary difference in the proposed approach is that both the marginal and the joint models appropriately incorporate unordered categorical variables along with count, binary, and continuous variables. The algorithm to create one synthetic observation is summarized below:

\begin{align*}
  & \tilde{y}_{cat} \sim multinomial(\hat{\boldsymbol{p}}) \\
 &\tilde{z}_{cat} \mid \tilde{y}_{cat}, -  \sim TMVN(\boldsymbol{\alpha_{cat}, C_{cat}, l,u}) \\
 &\tilde{z}_{-cat} \mid \tilde{z}_{cat}, - \sim MVN(\boldsymbol{\alpha^{*}, C^{*}}) \\
 & \tilde{y}_{-cat_{j}} = \hat{F}_{j}^{-1}\{\Phi(\tilde{z}_{-cat_{j}})\}\ 
\end{align*}
where $(\boldsymbol{\alpha_{cat},C_{cat}})$ is the sub-vector and sub-matrix of $(\boldsymbol{\alpha, C})$ corresponding to categorical components, $(\boldsymbol{l,u})$ are the vectors of lower and upper truncation limits defined in \eqref{trunc} and determined by the realization of $\tilde{y}_{cat}$, and 
$\boldsymbol{\alpha^{*}} = \boldsymbol{C}_{-cat,cat}\boldsymbol{C}_{cat,cat}^{-1}(\boldsymbol{\alpha}_{cat} -\tilde{z}_{cat})$ and  $\boldsymbol{C^{*}} = \boldsymbol{C}_{-cat,-cat} - \boldsymbol{C}_{-cat,cat}\boldsymbol{C}_{cat,cat}^{-1}\boldsymbol{C}_{cat,-cat}$ are the moments for the conditionally multivariate Gaussian distribution $z_{-cat} | z_{cat}, -$. Note that this algorithm uses posterior samples of the key parameters $(\boldsymbol{\alpha, C})$.

\section{Adaptations for Nonlinear Regression}\label{nonlinear}
 For our dataset, perhaps the most desired quality of the synthetic data is to maintain the relationships between educational development (EOG test scores) and social or environmental exposures. However, the Gaussian copula may be insufficient for modeling more complex relationships due to the linear factor model on latent $\boldsymbol{z}$. While a semiparametric, fully nonlinear joint model is an attractive alternative to consider, the required modeling and computational demands introduce significant challenges, including a high risk of over-fitting. In the case of synthetic data generation, this may result in high attribute disclosure risk -  the probability that an adversary could use synthetic data to infer specific sensitive information/features on real individuals \citep{hu2014disclosure,Hu2019}  -  and jeopardize privacy. Instead, we deploy the general flexibility of our framework specifically for modeling the EOG test scores conditional on the remaining variables using a semiparametric regression model. 
 
 Let $\boldsymbol{Y_{resp}}$ denote the response variables of interest and $\boldsymbol{Y_{cop}}$ the remaining variables. We decompose the data synthesis into $P(\boldsymbol{Y_{cop}}, \boldsymbol{Y_{resp}}) = P(\boldsymbol{Y_{cop}}) P(\boldsymbol{Y_{resp}} | \boldsymbol{Y_{cop}})$, where $P(\boldsymbol{Y_{cop}})$ is given by the extended rank-probit likelihood factor model and $P(\boldsymbol{Y_{resp}} | \boldsymbol{Y_{cop}})$ is a semiparametric regression model. By isolating these variables $\boldsymbol{Y_{resp}} $ from the joint model and synthesizing them conditional on social and environmental exposures using a flexible, semiparametric regression, we enhance the complexity of the synthetic data without jeopardizing the convenience of creating the vast majority of the synthetic dataset through the joint copula model.

Using this strategy, we develop a nonlinear regression model motivated by the semiparametric Gaussian copula, where the augmented data adhere to the rank-likelihood and observed data are realized by linking latent variables through the empirical CDF. By treating $\boldsymbol{Y_{cop}}$ as covariates, the conditional model for each $y_{resp_{ij}}$ given  $\boldsymbol{y_{cop_{i}}}$ is
\begin{align}
     z_{resp_{ij}}   &= f(\boldsymbol{y_{cop_{i}}}) + \epsilon_i, \quad \epsilon_i \stackrel{iid}{\sim} N(0, \sigma^{2}) \label{eq27}\\
    y_{resp_{ij}} &= \hat{F}^{-1}\{\Phi(z_{resp_{ij}} )\} \label{eq28}
\end{align}
where  $f$ is a flexible regression function. Since the EOG test scores are integer-valued, \eqref{eq27}-\eqref{eq28} presents a semiparametric version of \cite{kowal2020simultaneous}. More relevant, \eqref{eq27}-\eqref{eq28} is coupled with a rank likelihood for $\{z_{resp_{ij}} \}$, so these latent data preserve the ordering in the observed data $\{y_{resp_{ij}}\}$. The presence of the empirical CDF $\hat F$ ensures that the posterior predictive draws associated with \eqref{eq27}-\eqref{eq28} will maintain the marginal properties of $\{y_{resp_{ij}}\}$, while the regression function $f$ captures the conditional dependence of $\boldsymbol{Y_{resp_{j}}} | \boldsymbol{Y_{cop}}$. 

For the regression function $f$, we use a Bayesian Additive Regression Tree (BART), which is capable of modeling nonlinear and low-order interactions among the covariates $\boldsymbol{Y_{cop}}$ \citep{chipman2010bart}. Because of the conditional Gaussianity in \eqref{eq27}, the parameters of $f$ can be sampled using existing algorithms for Gaussian BART models, such as the \verb|dbarts| package in \verb|R|. Specifically, the sampling algorithm has two blocks:
\begin{enumerate}
\item \textbf{Sample the latent data:}  $z_{resp_{ij}} \mid f, \sigma \sim TN( f(\boldsymbol{y_{cop_{i}}}),\sigma^{2},z_{i}^{l}, z_{i}^{u})$ for $i=1,\ldots,n$, where $z_{resp_{ij}}^{l} = max\{z_{resp_{-ij}}: y_{resp_{-ij}}<y_{resp_{ij}}\}$, $z_{resp_{ij}}^{u} = min\{z_{resp_{-ij}}: y_{resp_{-ij}}>y_{resp_{ij}}\}$, and $z_{resp_{-ij}} = \boldsymbol{z_{resp_{j}}}\setminus z_{resp_{ij}}$
\item \textbf{Sample BART parameters:} $f,\sigma | \boldsymbol{z}$
\end{enumerate}
The first block follows from the rank likelihood construction, while the second block proceeds using a Gaussian BART sampling step. 

For data synthesis, let $\tilde y_{cop_i}$ denote a draw from the posterior predictive distribution of the copula model for $\boldsymbol{Y_{cop}}$. The nonlinear outcomes are synthesized as $\tilde y_{resp_i} = \hat F^{-1}\{\Phi(\tilde z_{resp_{ij}})\}$ where $\tilde{z}_{resp_{ij}}\sim N( \hat f(\tilde y_{cop_i}), \hat \sigma^2)$. In the sampling step for $\tilde {z}_{resp_{ij}}$, we use the posterior means $\hat f$ and $\hat \sigma$ of $f$ and $\sigma$, respectively. Although we could use any posterior samples of $f$ and $\sigma$ at this step, the use of the posterior means simplifies and stabilizes the data synthesis. Either way, we note that the nonlinearity afforded by \eqref{eq27}-\eqref{eq28} requires minimal computational burden: the other variables $\boldsymbol{Y_{cop}}$ are modeled and synthesized separately using the extended rank-probit likelihood and factor model, while the modeling and synthesis of $\boldsymbol{Y_{resp}} | \boldsymbol{Y_{cop}}$ features the same computational complexity as a BART probit regression.

\section{Results}\label{results}
\subsection{Simulation Study}\label{sims}
We first highlight the gains of the rank-probit likelihood (RPL) over the rank likelihood (RL; \citealp{hoff2007extending}) as a model for mixed unordered categorical and numerical data through a simulation study.  The proposed approach provides two main advantages over the RL: i) it provides a valid data generating process for unordered categorical data, allowing for the construction of viable synthetic datasets and ii) it allows ordinal variables with few levels to be treated as unordered categorical with improved performance.

First,  we construct a simulated dataset with two variables: one unordered categorical $(x_{1})$ with five levels and one integer-valued variable $(x_{2})$. The simulation design is comparable to the relationship between mother's race ($x_1$) and EOG test score ($x_2$) in the North Carolina dataset. We select the marginal proportions for each category of $x_1$,  $\boldsymbol{p}= (.10,.40,.15,.20,.15)$, and simulate $x_1$ from a multinomial distribution. Next, conditional on the level, we simulate the integer-valued variable  $x_{2} \mid x_{1} = l \sim \text{Poisson}(r_{l})$ where the Poisson rate $\boldsymbol{r} = (342,344,346,348,352)$ is specific to each category level. This process is repeated independently for each of $n = 5000$ individuals. 
Note that the levels of $x_{1}$ are assigned integer values from 1 and 5 for notational simplicity, but we do not assume an ordering among the levels of $x_1$.

For both the RPL and the RL, 500 synthetic datasets are generated, and the properties of these datasets are compared to those in the original data. For the RPL, $x_{1}$ is given the full binarization in \eqref{DO-Probit}, the copula correlation matrix and a non-zero intercept $\boldsymbol{\alpha}$ are estimated through the factor model \eqref{eq22}, and datasets are synthesized as in Section~\ref{synthesis}. 
For the RL, we employ the workaround described in Section~\ref{erpl}: the categorical variable is one-hot-encoded as 4 independent binary variables with base category $l=1$. Estimation of the copula proceeds under default prior specification using the \verb|sbgcop| package \citep{hoff2007extending}, and data synthesis is carried out through posterior predictive sampling. The MCMC for both models was run for 15,000 iterations, of which 9,000 were discarded as burn-in. Each of the 500 synthetic datasets has the same size ($n=5000$) as the original simulated data.

For each synthetic dataset, we recorded the mean of $x_{2}$ by categorical level, $\bar{X}^{syn}_{l}$ for $l=1,\ldots,5$, akin to computing the average EOG test score by mother's race level in the North Carolina dataset. We then compute the mean squared error between this value and the ground truth, $\mbox{MSE} = 5^{-1} \sum_{l = 1}^5 (\bar{X}^{syn}_{l} - \bar{X}_{l})^2$, where $\bar{X}_{l}$ is the sample mean of $x_{2} \mid x_{1} = l$. This MSE measures the ability of the synthetic data to capture the dependence between an unordered categorical variable and an integer-valued variable. We compute the average and standard error of this statistic across all 500 synthetic datasets. In addition, we record the proportion of individuals for which the categorical variable $x_1$ is erroneously assigned multiple categories. A sizeable proportion of multiple classified individuals in a synthetic dataset is problematic since these individuals would have to be discarded to maintain consistent categorizations with the confidential data. Such an \emph{ad hoc} process can skew the marginal categorical proportions and negatively impact other bivariate or joint dependencies, and requires further modifications to ensure that the synthetic data have the same dimensions $(n)$ as the original data. 

The results are presented in Table~\ref{sim-comp}. Clearly, the proposed RPL approach is better suited for modeling relationships between categorical and numerical measurements. Conditional distributions $x_{2} \mid x_{1} = l$ consistently and closely concentrate around ground truth group means. In addition, our method for data synthesis provides universally feasible categorical observations, while the RL with one-hot-encoding generates infeasible categorical data in more than 12\% of observations.

\begin{table}[H]
\begin{center}
\begin{tabular}{l|c|c|c}
    Method & Avg. MSE & S.E. MSE &$\%$ Multiple Classified Individuals  \\\hline
     RPL& \textbf{0.451} &  \textbf{0.324} & \textbf{0} \\ \hline
     RL & 2.359 &0.763& 12.4
 
\end{tabular}
\end{center}
\caption{The proposed RPL more closely and consistently preserves categorical-numeric relationships and completely avoids erroneous assignment of multiple category levels for an individual. \label{sim-comp}}
\end{table} \vspace{-.5cm}

Further advantages of the RPL over the RL are apparent in the modeling of ordered categorical variables with few levels.  While the RL can only model such variables as ordinal, the RPL provides the option to model these variables as (unordered) categorical variables. Perhaps surprisingly, the latter approach can offer significant improvements, which motivates us to model mother's education level ($k=4$) as (unordered) categorical in the RPL. 

We return to the same simulated dataset used in the previous demonstration, but this time, the integer assignments to $x_{1}$ are treated as numerical values in the RL, which is how \cite{hoff2007extending} and \cite{murray2013bayesian} propose to handle ordinal variables.  In this setting, $x_{1}$ now resembles the mother's education variable in the NC data set, and we study how distinguishing it as ordinal or categorical in the copula model may affect the resulting synthetic data.

 Since the expectation of $x_{2} \mid x_{1} = l$ is increasing in $l$, an ordinal relationship would appear to be reasonable. Yet for the RPL, we continue to treat this variable as categorical. We once again create 500 datasets under each method, and record group conditional means $\bar{X}_{l}^{syn}$ for each. The distribution of these group conditional means across 500 synthetic datasets, along with overlayed $95 \%$ intervals (dotted green lines), and the ground truth group conditional mean $\bar{X}_{l}$ (solid red line) is shown in Figure 1.

\begin{figure}[H]
    \centering
    \includegraphics[width = .75\linewidth, keepaspectratio]{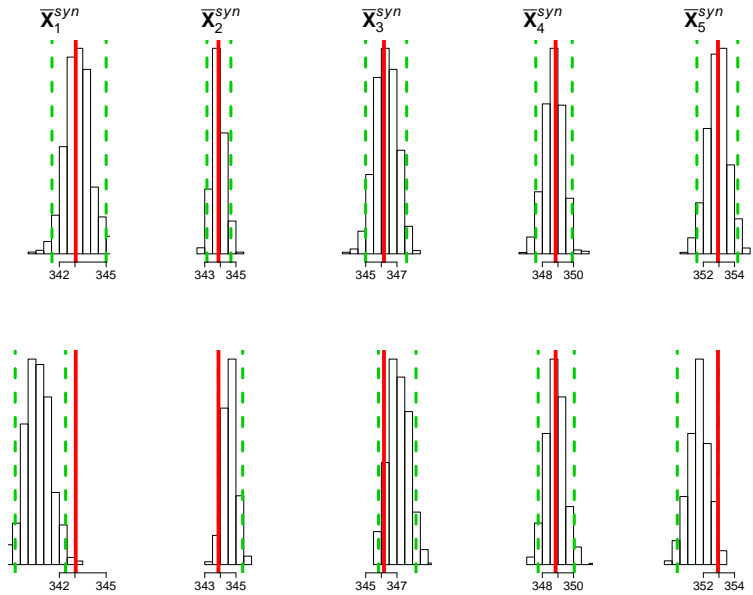}
    \caption{Distribution of $\bar{X}^{syn}_{l}$ across synthetic datasets where $x_{1}$ is treated as categorical (top row) or ordinal (bottom row). The performance of the RPL in modeling ordinal variables with few levels as categorical suggests additional advantages over the RL.}
    \label{sim-study}
\end{figure} \vspace{-.5cm}

When the ordering on $x_{1}$ is acknowledged and the RL is employed, synthetic data are decidedly inconsistent; the distribution of $\bar{X}^{syn}_{l}$ across synthetic datasets concentrates far away from the ground truth for all but one level. Under the RPL, the analyst can choose whether to treat such variables as ordinal or categorical, and this choice can greatly improve the quality of the synthetic data.  We provide further simulation results in the supplementary materials regarding the sensitivity of these results to the number of the levels of ordinal $x_{1}$ \cite{supp}. Based on this analysis, we recommend that any ordinal variable with fewer than 10 levels be treated as categorical within the RPL. 

\subsection{Synthetic North Carolina Data}\label{sub-results}
 We now turn to the creation of  synthetic North Carolina data. In the process, we compare the performance of our approach  in providing consistent regression inference with reference to the confidential set  to its Bayesian competitor (RL) and the popular nonparametric alternative, classification and regression trees (CART).  We also study the utility-attribute disclosure risk trade-off of publicly releasing fully synthetic data under each method.  
 
 The dataset, described in Table~\ref{datad}, contains $n = 19,364$ records of $p=23$ variables. Both mother's race and mother's education are modeled as unordered categorical variables. Although a natural ordering exists for mother's education, the simulations from Section~\ref{sims} strongly suggest that because the variable has so few levels (four), accuracy can be improved significantly by treating this variable as an unordered categorical variable. In addition, we target the synthesis for prediction of EOG reading and mathematics scores based on the remaining variables,  since the focal point of study on the NC dataset is understanding the potentially complex relationships between these educational outcomes and demographic and health information, environmental exposures, and social stressors.

Specifically, we apply the model from Section~\ref{nonlinear} for EOG reading and mathematics scores as the response variables $\boldsymbol{Y_{resp}}$. By using the BART model in \eqref{eq27}-\eqref{eq28}, our data synthesis can capture and reproduce nonlinear and interactive associations among these demographic, health, and exposure variables for predicting  cognitive development. The MCMC algorithm for the remaining variables $\boldsymbol{Y_{cop}}$, modeled using the extended rank-probit likelihood and factor model, was run for 50,000 iterations, of which 25,000 were discarded as burn-in. The sampler for $\boldsymbol{Y_{resp}}$ was run for 1,100 iterations, of which 100 were discarded as burn-in.
\subsection{  Basic Utility Properties}
\begin{figure}[H]
    \centering
    \includegraphics[width=0.5\textwidth, keepaspectratio]{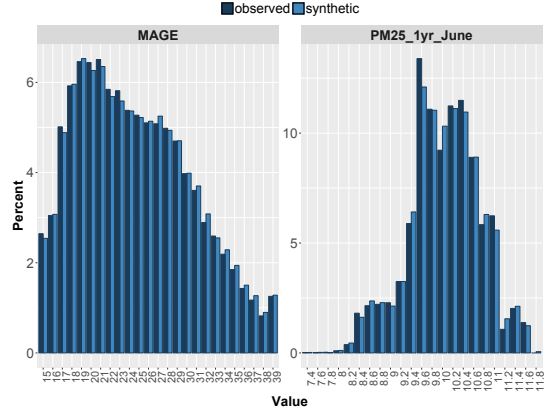}
    \caption{Comparison of synthetic (light blue) and observed (dark blue) marginal distributions for mother's age (left) and chronic PM 2.5 exposure (right). Despite the notable non-Gaussianity, the synthetic data distributions closely match the confidential data distributions. \label{fig-marg}}
\end{figure}

 We first demonstrate simple properties of synthetic data sets produced under our proposed framework. This allows us to highlight the advantages of our semi-parametric copula model in modeling categorical and numerical measurements jointly. We first conduct univariate (Figure~\ref{fig-marg}) and bivariate (Table~\ref{tab-biv}) assessments of data utility by comparing synthetic and confidential data.  Because our method utilizes empirical CDFs for continuous (with an additional kernel smoother) and count variables, the synthetic data capture notably non-Gaussian margins. In addition, we see consistent cross-tabulations of the two categorical variables, mother's education and mother's race, between synthetic and confidential data based on our approach to data synthesis outlined in Section~\ref{synthesis}. 
\begin{table}[H]
\centering
\textbf{Synthetic (Observed) Cross-Tabulations of Mother's Education and Race}
\begin{tabular}{l|r|r|r|r|r}

  & NH White & NH Black & Hisp & NH Asian/PI & NH Other\\

\hline

No High School Diploma & 0.091(0.093) & 0.086(0.085) & 0.087(0.091) & 0.005(0.005) & 0.002(0.003)\\

\hline

High School Diploma & 0.172(0.167)& 0.134(0.135) & 0.032(0.032) & 0.006(0.007) & 0.002(0.002)\\

\hline

Some College/ Associates Deg. & 0.118(0.116) & 0.078(0.078) & 0.008(0.008) & 0.003(0.003)& 0.002(0.001)\\

\hline

Bachelor's or Above & 0.126(0.127) & 0.037(0.036) & 0.004(0.004) & 0.005(0.005) & 0.001(0.001)\\

\hline

\end{tabular}
\caption{The synthetic RPL data preserve multivariate categorical properties observed in the original data. \label{tab-biv}}
\end{table}

\vspace{-.5cm}

\vspace{-.5cm}

\subsection{Comparing Regression Utility}

Regression of EOG math and reading scores on social and environmental exposures can be used to uncover main and low order interaction effects relating exposure profiles of children to educational development \citep{BVS}.  An important assurance of the quality of our synthetic data is whether regression models fit to synthetic and confidential data produce highly similar results. This type of utility must be prioritized in the creation of our synthetic dataset given how it will likely be analyzed by others.

To assess these conditional or regression associations, we fit Bayesian linear regression models to both the synthetic and confidential data and compare the inference on the resulting regression coefficients. EOG reading scores are regressed on social stressors, environmental exposures, and other baseline demographic and health information. In addition, we include several key interactions between social and environmental exposures as in \cite{BVS}, creating what we subsequently refer to as the augmented linear model. In both models, numerical predictors are scaled to mean zero and unit variance. The regression coefficients are assigned horseshoe priors \citep{carvalho2010horseshoe} so that in addition to comparing the strength of posterior signals, we can also evaluate whether synthetic data produces consistent parameter shrinkage. This procedure applied to the synthetic data is a posterior predictive \emph{functional} that is specifically customized for our task---reliable data synthesis for regression---as advocated by \cite{kowal2021fast}.

 We design a study to compare our framework with RL and CART.  For this study, we simulate five synthetic datasets under each method, and use the combining rules for inference on partially synthetic data from \cite{Reiter2003} to derive point estimates and confidence intervals for regression coefficients. By comparing these estimates to those derived from the confidential dataset, we are able to understand the extent to which each synthesis model produces synthetic data that provides consistent regression inference. 

We adopt two specific measures of synthetic data utility from \cite{snoke2018general} to conduct these comparisons: 95\% confidence interval overlap ($CIO$), and standardized coefficient mean-squared error ($MSE$). These statistics provide an evaluation with reference to the confidential dataset of two desired qualities in regression inference on synthetic data: consistency in uncertainty quantification and point estimation.

We compute the $CIO$ for the $j$-th regression cofficient as
\[CIO_{j} = 0.5\left(\frac{min(u_{obs},u_{syn}) - max(l_{obs},l_{syn})}{u_{obs} - l_{obs}} + \frac{min(u_{obs},u_{syn}) - max(l_{obs},l_{syn})}{u_{syn} - l_{syn}} \right)\]
where $(l_{obs},u_{obs})$ and $(l_{syn},u_{syn})$ are the lower and upper endpoints of the 95\% uncertainty interval for coefficient $j$ estimated from the confidential dataset and the pooled synthetic datasets respectively. Higher values of $CIO_{j}$ indicate that the uncertainty quantification for regression coefficient $j$ is similar between synthetic and confidential data.

To compute the $MSE_{j}$ for each predictor, we take the posterior mean and standard deviation of the samples produced from the fit to the confidential dataset, $\hat{\beta}_{j_{obs}}$ and $\hat{\sigma}_{\beta_{j_{obs}}}$, and the pooled point estimate $\hat{\beta}_{j_{syn}}$, which is the average of the posterior means across the five synthetic datasets, and compute $( \hat{\beta}_{j_{obs}} - \hat{\beta}_{j_{syn}})^{2}/\hat{\sigma}^{2}_{\beta_{j_{obs}}}$. This statistic penalizes large deviations between point estimates on confidential and synthetic data for which there is low posterior uncertainty in the confidential data. Therefore, synthetic data with low MSEs reflect a pattern of consistent point estimation.

For the RPL synthetic datasets, we use the posterior predictive sampling algorithms outlined in Sections ~\ref{synthesis} and ~\ref{nonlinear} using posterior samples from the copula and BART models fit to the confidential dataset. For the RL synthetic datasets, we estimate a Gaussian copula on the confidential data under default settings in the \verb|sbgcop| package. We employ the categorical workaround for mother's race, encoding the variable as 4 indicators with non-hispanic white as the base level,  and treat mother's education as an ordinal variable with increasing levels from 1 (No High School Diploma) to 4 (Bachelor's or Above). Posterior predictive sampling is once again used to create synthetic datasets, and multiple classified individuals are discarded since synthetic datasets should adhere to the same categorization conventions as the confidential data.

For CART, we employ a tuning strategy to select an ordering for synthesis.  As mentioned in Section~\ref{intro}, the primary limitation to sequential synthesis is the need to declare an ordering of the variables for modeling.  Ideally, the analyst should choose the ordering that provides the most useful synthetic data.  To accomplish this, we first create 100 synthetic datasets with unique orderings using default settings for CART in the \verb|synthpop| package in \verb|R|. Based on our utility criteria of consistent regression inference, only the last two variables in each ordering, EOG reading and math scores, are fixed so that the synthesis model is supplied with the most conditional information. The ordering of the remaining 21 variables are randomly selected.

For each of the 100 synthetic CART datasets, we fit the augmented linear model and record $\overline{CIO} =\frac{1}{p}\sum_{j=1}^{p}CIO_{j}$ and $\overline{MSE} = \frac{1}{p}\sum_{j=1}^{p}MSE_{j}$. In addition, we record the propensity score mean-squared error ($pMSE$), which is a  measure of general synthetic data utility \citep{snoke2018general}. To compute the $pMSE$ for a synthetic dataset, synthetic and confidential observations are pooled, and an indicator as to whether each observation in the combined dataset is synthesized or real is added.  A logistic model is fit, where the indicator is regressed on all of the other variables in the dataset. $pMSE$ averages how well the model can discriminate real from synthetic data points, relative to a random guess. A low value of this statistic indicates synthetic data are difficult to distinguish from confidential data, indicating high overall synthetic data similarity.

To choose a single CART ordering among the 100 considered, we average the metrics previously introduced into a cohesive summary of a synthetic data set's utility, as was done in \cite{taub2021synthetic}.  Specifically, for each CART synthesis, we compute the aggregated utility, $U$, as 
\[U = \frac{\overline{CIO} + (1-\overline{MSE}) + (1-4*pMSE)}{3}\]
This statistic balances regression and general utility, taking into account the scale of each statistic. Synthetic data with perfect regression utility would have $\overline{CIO} =1$ and $\overline{MSE} = 0$, reflecting matching point estimates and uncertainty quantification. Similarly, synthetic data that is globally true to the confidential dataset would have a $pMSE$ very close to 0. As such, the statistic is constructed to such that its maximum possible value is 1, with higher values indicating more useful synthetic data. The CART ordering that produced the highest $U$ was selected for the subsequent analyses.

For each method, we repeat the following process 20 times: i) synthesize 5 datasets ii) pool the datasets to derive point and interval estimates for regression coefficients and iii) compute $CIO_{j}$ and $MSE_{j}$ for each coefficient. To compare the RPL to its competitors, the difference in these statistics by coefficient allows us to evaluate whether one method out-performs the other. For instance, to evaluate whether the RPL provides more similar uncertainty quantification than CART for coefficient $j$, we would like to estimate the distribution of $CIO^{RPL}_{j} - CIO^{CART}_{j}$. We rely on bootstrap sampling in order to do so; we repeatedly and randomly sample (with replacement) $CIO^{RPL}_{j}$ and $CIO^{CART}_{j}$ from the 20 replicates, and compute the difference.

\begin{figure}[H]
    \centering
        \includegraphics[width = .49\textwidth, keepaspectratio]{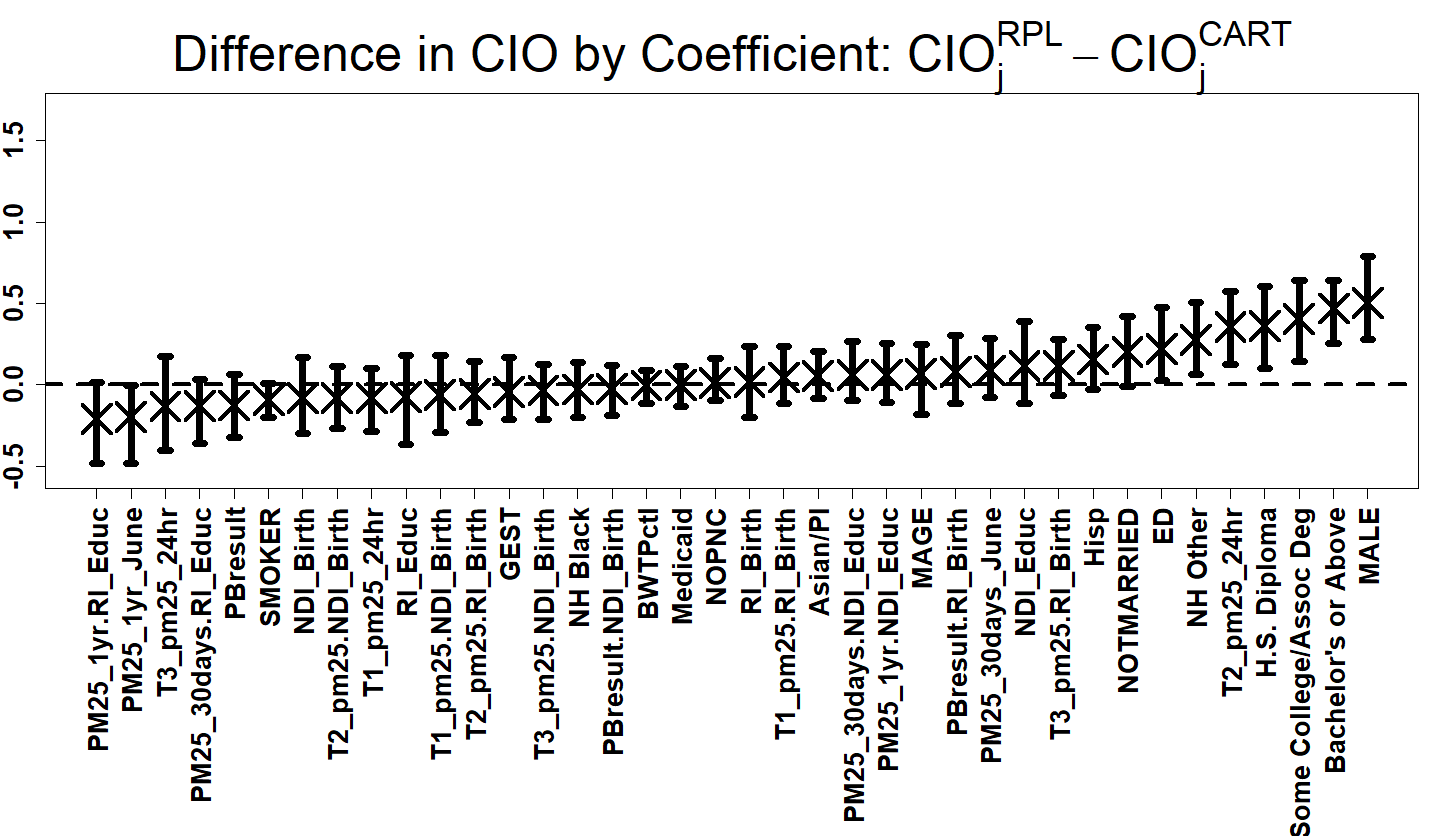}
        \includegraphics[width = .49\textwidth, keepaspectratio]{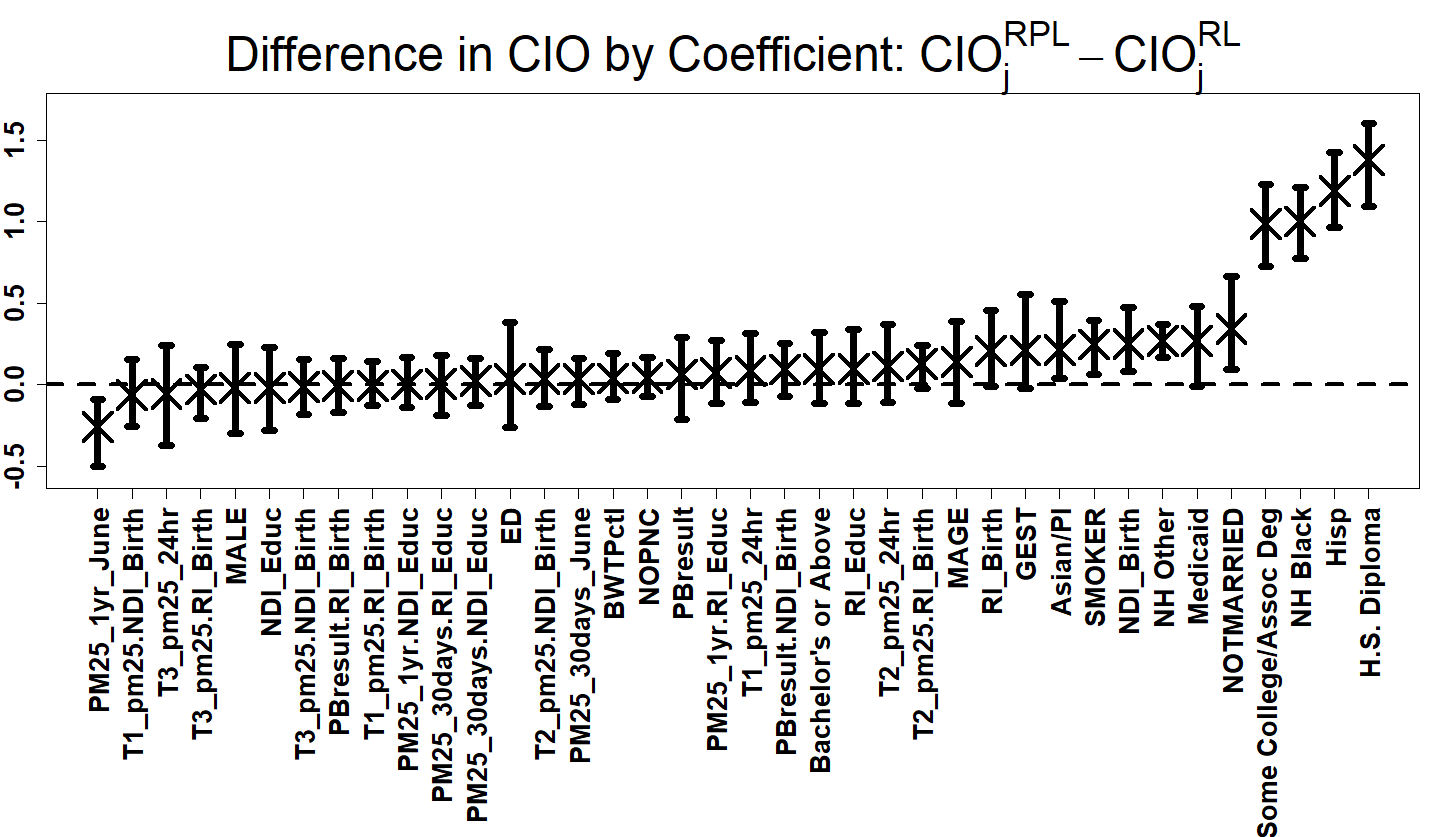}
        \includegraphics[width = .49\textwidth, keepaspectratio]{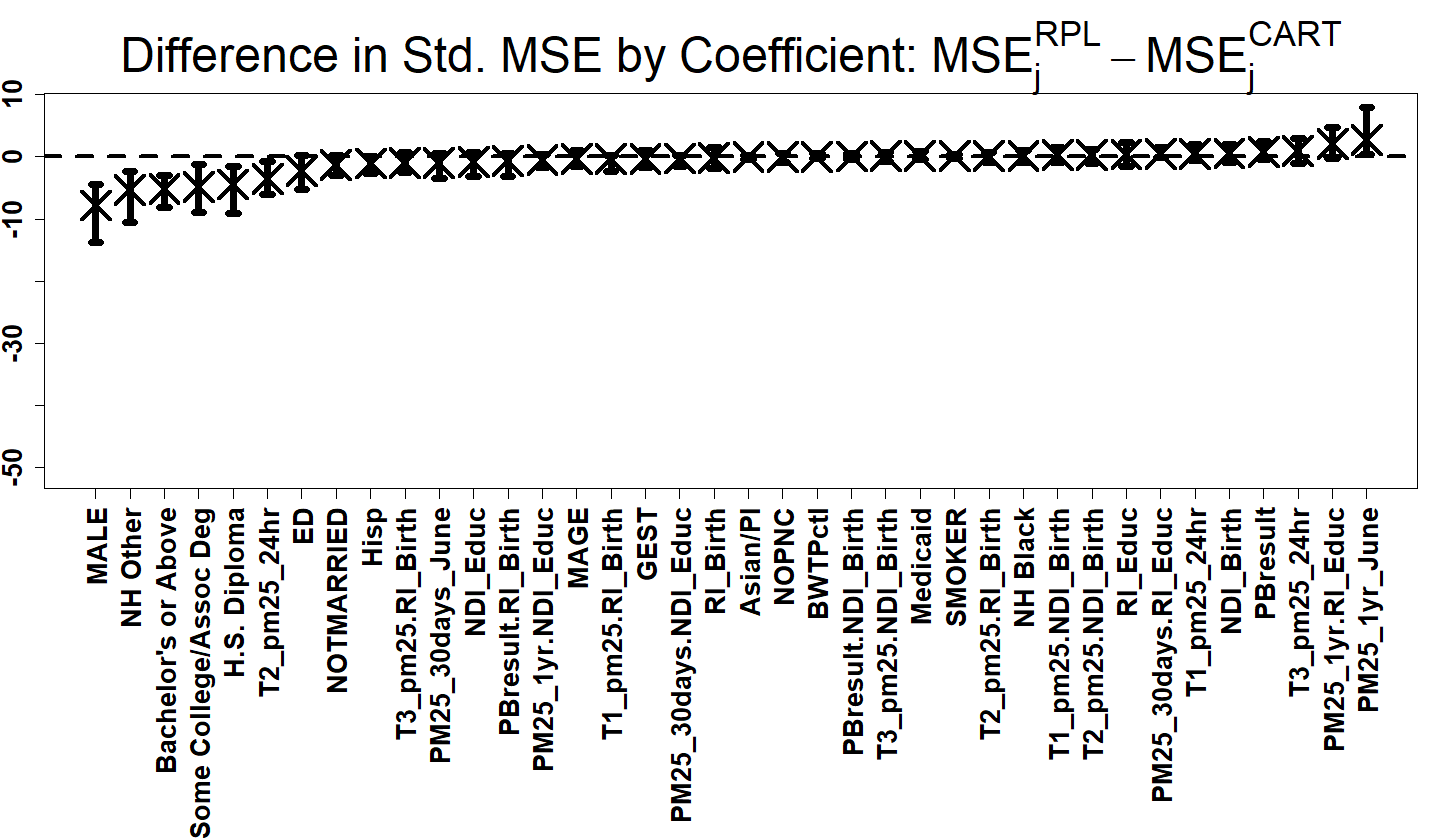}
        \includegraphics[width = .49\textwidth, keepaspectratio]{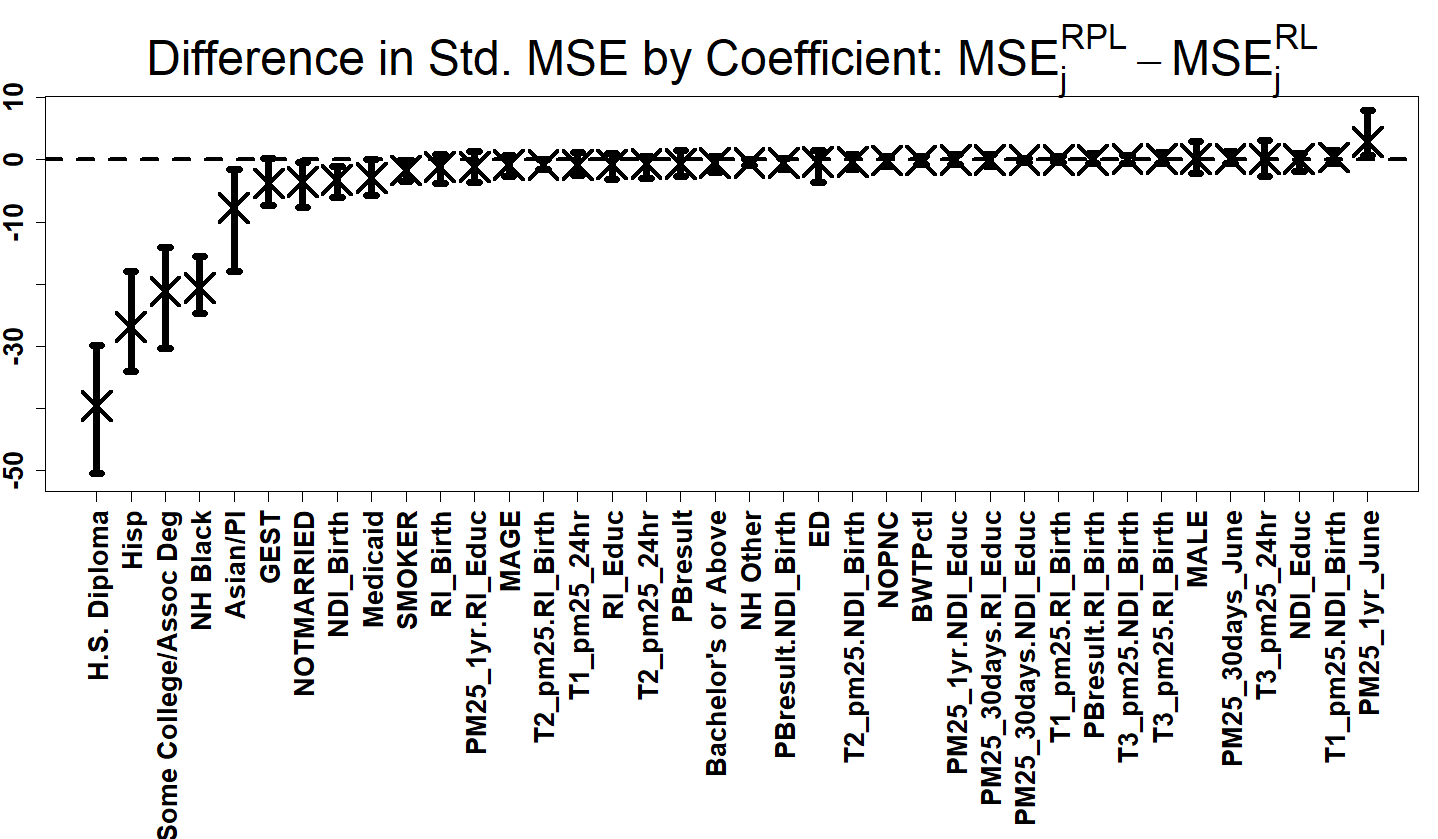}
    \caption{Mean and 95\% uncertainty intervals for the difference in specific utility between RPL and its competitors across 20 repetitions of simulating $m = 5$ synthetic datasets.  Along the rows, we vary the utility statistic, while the columns compare RPL to CART (first column) and RL (second column). The results demonstrate that the RPL provides substantial improvement to the consistency of regression inference over its Bayesian competitor, and significant improvement over CART.}
    \label{fig-coef}
\end{figure}

Figure~\ref{fig-coef} plots the means and 95\% uncertainty intervals of 1000 bootstrap samples for the differences in $CIO$ and $MSE$ between RPL and its competitors.  By construction, these intervals serve to evaluate our method: any interval that does not contain zero provides significant evidence that one method outperforms the other. Negative differences for $MSE$ indicate that point estimation is more faithful to the confidential dataset for our method, while positive differences for $CIO$ show our method produces more similar uncertainty quantification relative to other method used to estimate this distribution. Overlaid onto the plot is a horizontal line at 0, which provides this reference point.

The results demonstrate that for several regression coefficients, the RPL provides significant improvement in the consistency of both uncertainty quantification and point estimation. The results are most striking in the comparison between the RPL and its Bayesian competitor, the RL; the right-most coefficients for $CIO$ and left-most coefficients for $MSE$ highlight improvements for our method, as the credible intervals entirely and substantially miss the zero line. As might be expected, the substantial improvements under our method come for the variables that are better modeled as unordered categorical under the RPL, Mother's Education and Mother's Race Group.

In comparison to CART the gains are more modest. However, the RPL provides significantly more consistent uncertainty quantification and point estimation for 7 predictors, whereas such a statement can only be made about one predictor, \verb|PM25_1yr_June|, for CART.

These results highlight the effectiveness of the targeted synthesis strategy and the ability of our model to seamlessly incorporate unordered categorical variables. When a specific regression model is a focal point of study in the confidential data, it is beneficial to pay special attention to the synthesis of its response variables. This underscores the importance when synthesizing a confidential dataset of i) identifying inferential models of interest and their likely response variables to inform a synthesis order ii) using a suitably powerful synthesis model for response variables of interest. 

\subsection{The Utility-Risk trade-off}
In order to make synthetic data publicly available, the privacy implications of its release must be considered. For fully synthetic data, the risk of re-identification is generally considered non-existent, since synthetic individuals do not correspond to any individuals in the confidential dataset. However, the risk of sensitive attribute disclosure may persist, since an adversary may use synthetic data to accurately identify specific sensitive attributes of real individuals. \cite{reiter2014bayesian} formulate a Bayesian estimation procedure for attribute disclosure risk. Though this computation would be possible for RPL and RL synthetic datasets, it requires a posterior distribution for synthesis model parameters, which CART lacks. We utilize alternative methods for calculation, but it should be mentioned that synthesis under the Bayesian paradigm is a major advantage of this work, as it provides the opportunity to more cohesively measure attribute disclosure risks.

Upon synthesizing many datasets,  a common practice for a data disseminating agency is to choose $m>1$ synthetic datasets for public release \citep{reiter2005releasing}. The selection of which datasets to release is guided by the natural desire for these synthetic data to be simultaneously useful and sufficiently private. However, the more useful the synthetic data, the more it  may  resemble the confidential data, and a trade-off is expected.  \cite{Duncan2001} study this trade-off as a function of the disclosure limitation procedure. Take the additive noise mechanism, which has been commonly used to de-identify confidential data. Adding more noise to the confidential data will simultaneously increase protectiveness and decrease utility. With a threshold for data utility, one can naturally study the maximum amount of noise that can be added to the confidential data such that it remains suitably useful.

By comparison, fully synthetic data simulated from a generative model is a fundamentally different procedure for maintaining data privacy. Often, it is difficult to deliberately increase the privacy of synthetic datasets generated from such models through a tuning parameter. Though CART may provide an exception to this statement -- one could tune regression/classification tree parameters such that synthesis models are over fit to the data resulting in a loss of privacy -- for our Bayesian synthesis model, there is no immediate mechanism through which we could deliberately increase the privacy of the resultant synthetic data and study the impact on utility. As a result, it is not clear how a utility-risk trade-off may emerge in the \textit{generation} of fully synthetic data.

Fortunately, this does not preclude the investigation and comparison of the utility attribute disclosure risk trade-off among the synthesis methods considered in this paper. However, we consider attribute disclosure risk as a function of several \textit{post-hoc} data dissemination procedures, as opposed to a function of the synthesizing mechanism. Prior to explaining these relationships, we must first highlight how we measure attribute disclosure risk.

For this work, we measure the attribute disclosure risk for EOG math scores. For both RPL and CART datasets, EOG math scores are synthesized last, which provides the synthesis model with the most conditional information. If the synthesizer suffers from overfitting, it may near perfectly reproduce observations in the confidential dataset, resulting in high disclosure risks.  To measure this risk, we extend the differential correct attribution probability (DCAP) metric of \cite{taub2018differential}, which measures the probability that an adversary correctly guesses a categorical sensitive target variable of an individual in the confidential dataset to accommodate count target variables. 

For calculation of DCAP, an intruder is interested in uncovering a particular sensitive attribute of an individual, and we assume that the intruder has access to a vector of information for each individual in the confidential set. We  define our confidential dataset $\boldsymbol{d_{o}} = \{\boldsymbol{K_{o}}, T_{o}\}$ in terms of the vectors of known ($\boldsymbol{K_{o}}$) information and the target variable(s) ($T_{o}$).   Similarly, a synthetic dataset can be partitioned $\boldsymbol{d_{s}} = \{\boldsymbol{K_{s}}, T_{s}\}$. For categorical targets, the correct attribution probability (CAP) for record $j$ in the observed dataset based on corresponding synthetic dataset $\boldsymbol{d_{s}}$ is calculated as
\begin{equation}\label{CAP}
    CAP_{sj} = \frac{\sum_{i=1}^{n} \left[T_{si} = T_{oj} \land \boldsymbol{K_{oj}} = \boldsymbol{K_{si}}\right]}{\sum_{i=1}^{n} \left[\boldsymbol{K_{oj}} = \boldsymbol{K_{si}}\right]} 
\end{equation}
where the square brackets are Iverson brackets and $n$ is the number of observations in the synthetic dataset. 

With categorical targets, a potentially fruitful strategy sees the adversary guess the mode among the targets corresponding to matches. For count-valued targets with a wider range of values, there often may not be a unique mode when employing this strategy.  Alternatively, we assume that the adversary would pool matches across the $m$ synthetic datasets released, and take the median of the synthetic targets in the match set as her guess for $T_{oj}$ \citep{taub2018differential}.
\begin{equation}\label{matchset}
    Med_{sj} = median\{T^{l}_{si}: \boldsymbol{K_{oj}} = \boldsymbol{K_{si}^{l}}, i = 1, \dots, n, \ l = 1, \dots, m\}
\end{equation}

Using this statistic, we develop the correction median attribution probability (CMAP) for confidential individual $j$ on synthetic data sets $ D = (d_{1}, \dots, d_{m})$ as
\begin{equation}\label{CAP}
    CMAP_{sj}^{\epsilon} =  \left[\lvert Med_{sj} - T_{oj}\rvert \leq \epsilon \right] 
\end{equation}
which indicates whether the adversary could use the matching strategy outlined to uncover confidential individual $j$'s target attribute, within some bound $\epsilon$ of the true value. If this is the case, we deem confidential record $j$ at-risk for attribute disclosure on the target variable. If there are no matches for $\boldsymbol{K_{oj}}$ in $D$, we take $CMAP_{sj}^{\epsilon} = 0$. To understand the disclosure risk of releasing $D$, we compute $\overline{CMAP_{s}^{\epsilon}} = \frac{1}{n}\sum_{j =1}^{n}CMAP_{sj}^{\epsilon}$, which is the proportion of confidential records at-risk using this median matching. 

This statistic can also be calculated with reference to confidential set ($\overline{CMAP_{o}^{\epsilon}}$), by  replacing $\boldsymbol{K_{si}}$ with $\boldsymbol{K_{oi}}$, and $T_{si}$ with $T_{oi}$. This value is important for understanding comparative disclosure risks, providing a view of the \textit{differential} correct median attribution probability. Though extremely conservative, comparisons to this quantity provide a view of the relative increase in privacy protection under more realistic assumptions on the intruder's prior knowledge of the confidential dataset.

\newpage
Armed with our attribute disclosure risk metric, we first study how it  varies across synthesis methods  as a function of three assumptions or decisions that the data disseminating agency must make when considering the public release of fully synthetic datasets:

\begin{itemize}
    \item \textbf{The number $\boldsymbol{m}$ of synthetic datasets to be publicly released}: We consider the release of $m = 5,10,20$ synthetic datasets. 
    \item \textbf{The amount of known prior information that the adversary is assumed to have ($\boldsymbol{K_{o}}$) about the confidential dataset}: We consider situations where the adversary has knowledge of $p = 4,5,6, \ \text{or} \ 7$ attributes from the set $\boldsymbol{K_{o}} = \{\text{Mothers Age}, \text{Mother's Education}, \\ \text{Mother's Race},\text{Gender},\text{Smoker}, \text{PBresult}, \text{Economically Disadvantaged}\}$, with $\boldsymbol{K_{o_{p}}} =\{k_{o_{1}}, \dots, k_{o_{p}}\}, p = 4,5,6,7$. 
    \item \textbf{The size of $\epsilon$, or the slack, for which we consider target $\boldsymbol{T_{oj}}$ at-risk in the calculation of $\boldsymbol{CMAP_{sj} =  \left[\lvert Med_{sj} - T_{oj}\rvert \leq \epsilon \right]}$}: For count-valued EOG math scores as the target, we consider $\epsilon = 0,1,2$ as reasonable values for which the median matching may produce an at-risk confidential record.

\end{itemize}

 \cite{reiter2009estimating} showed that increasing $m$ resulted in modest increases to attribute disclosure risks for partially synthetic data, while \cite{elliot2015final} found that increasing the amount of known prior information of the adversary decreases disclosure risks for fully synthetic data. We study these relationships  across synthesis methods for fully synthetic North Carolina data, changing the criteria for which we consider a confidential record at-risk through the slack parameter.

 We frame this analysis under a common data dissemination practice; usually, the statistical agency will generate and release $m$ synthetic datasets. To estimate the disclosure risk of a given synthesis method under this practice, we repeatedly (for 100 iterations) and randomly select $m$ of the 100 synthetic datasets synthesized from the previous section, and summarise the disclosure risks by method while varying the dissemination parameters. 
 
\begin{figure}[H]
\includegraphics[width = 9.5cm, keepaspectratio]{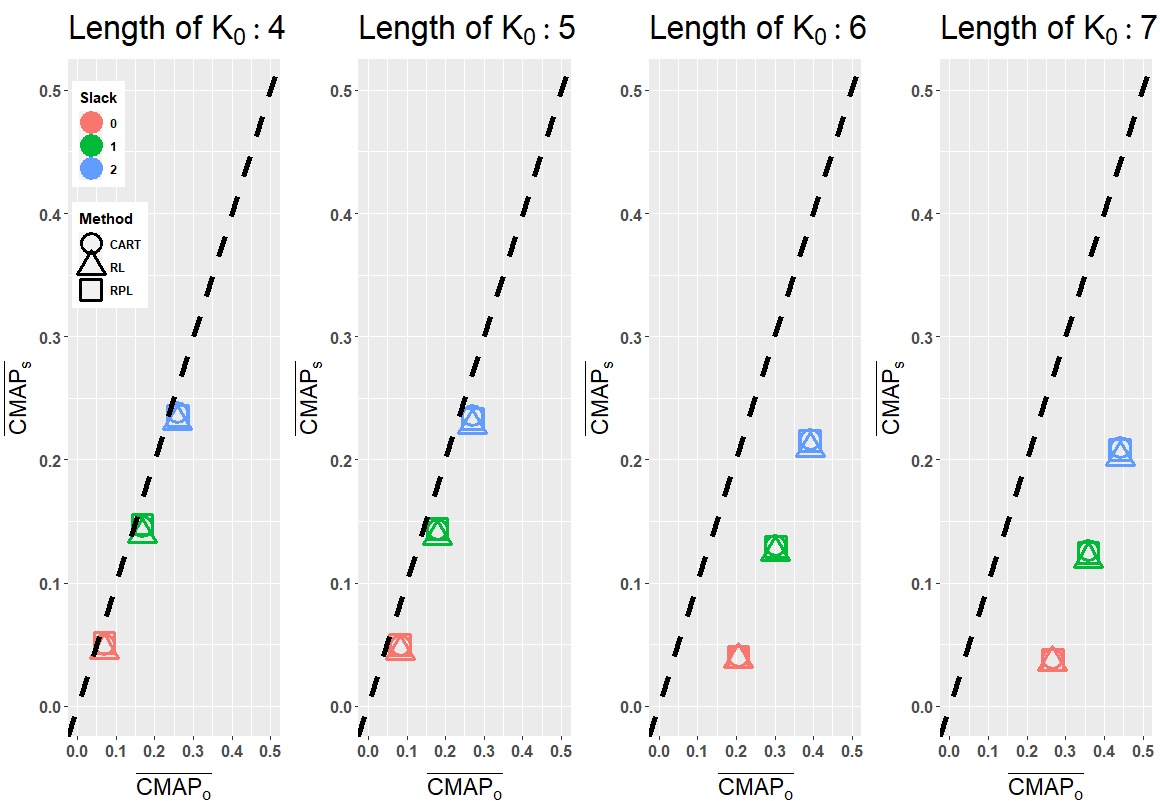}
\caption{ $m = 5$}
 \label{Risk}
 \end{figure}

\begin{figure}[H]
\includegraphics[width = 9.5cm, keepaspectratio]{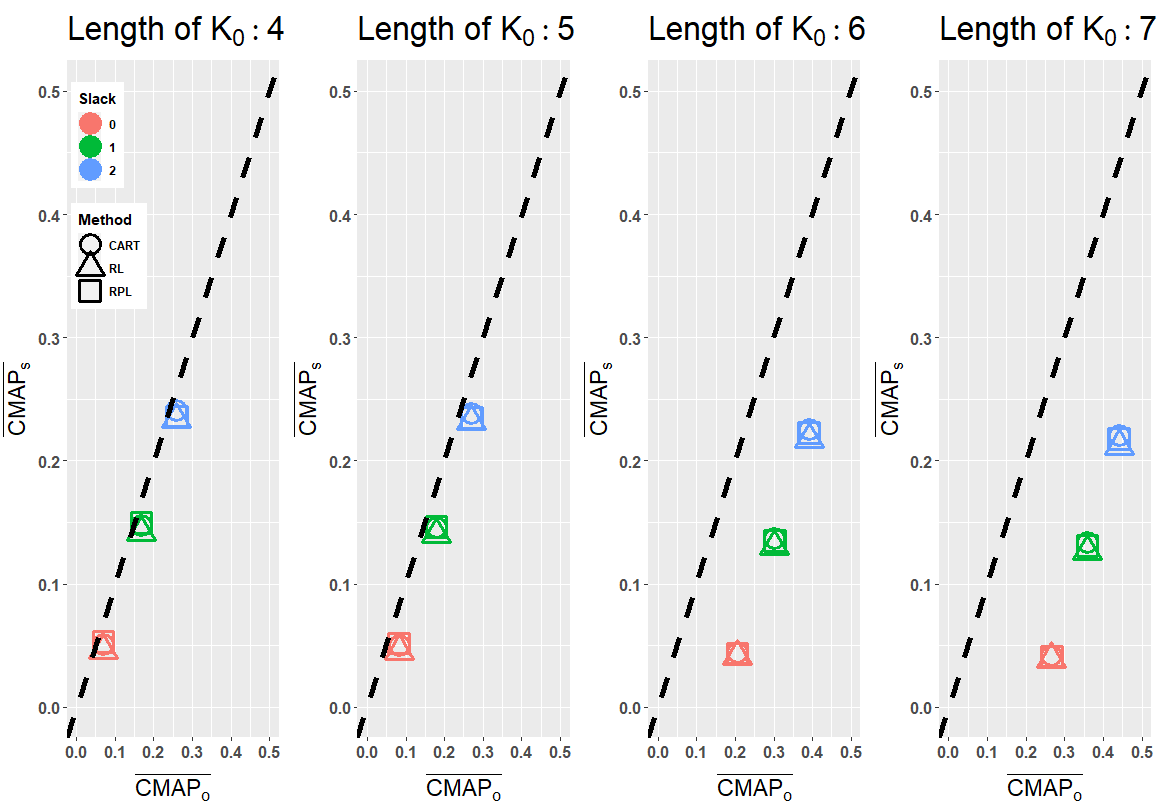}
    \caption{$m = 10$}
    \label{Risk1}
\includegraphics[width = 9.5cm, keepaspectratio]{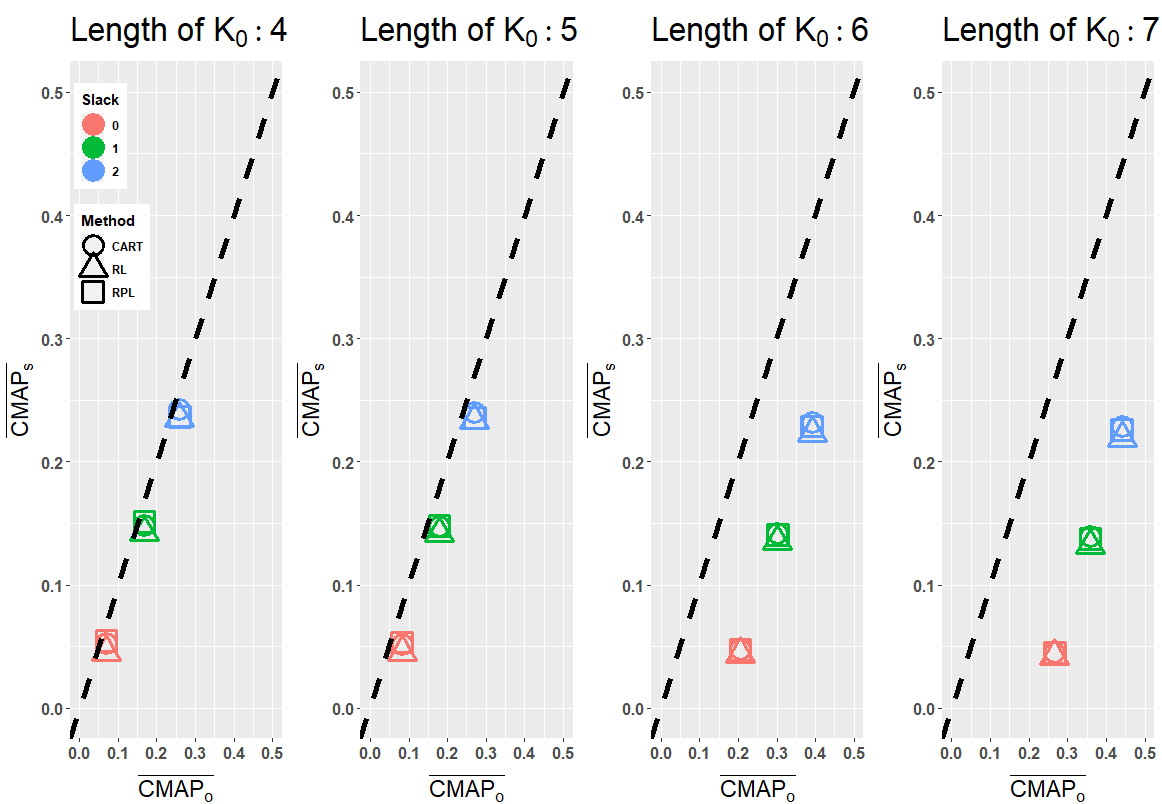}
    \caption{$m = 20$}
    \label{Risk2}
\end{figure}

 Figures~\ref{Risk}-\ref{Risk2} plot the average of $\overline{CMAP_{s}^{\epsilon}}$ across the 100 bootstrapped samples against $\overline{CMAP_{o}^{\epsilon}}$ for $m = 5,10, 20$, varying $\epsilon$ and the length of $\boldsymbol{K_{o}}$. Overlayed onto the plot is the line $y = x$, which provides an understanding of the reduction in attribute disclosure risk that the synthesis provides over the baseline. It should be noted here that $\overline{CMAP_{o}^{\epsilon}}$ remains constant across methods for each combination of $\boldsymbol{K_{o}}$ and $\epsilon$. In addition, the statistic does not depend on $m$, as it is calculated with reference to the confidential set.
 
\vspace{5mm}

As expected, $\overline{CMAP_{s}^{\epsilon}}$ is highly correlated with slack; for each $\boldsymbol{K_{o}}$, the disclosure risk increases approximately linearly in the amount of slack for which a confidential record is considered at-risk. Looking across the rows in each panel, $\overline{CMAP_{s}^{\epsilon}}$ slightly decreases for each slack value, consistent with the findings of \cite{elliot2015final}. This highlights the robust protectiveness of fully synthetic data, especially when the target is count-valued and an adversarial matching strategy is used.   Perhaps more importantly, as evidenced by the clustering of shapes in each panel, none of the synthesizers is clearly preferable in terms of its attribute disclosure risks regardless of slack or the length of $\boldsymbol{K_{o}}$. 

To compare the utility-risk trade-off across synthesis methods, we first compute the differential risk reduction for each combination of dissemination parameters, $\overline{CMAP_{o}^{\epsilon}} - \overline{CMAP_{s}^{\epsilon}}$. This statistic measures the privacy gain from the synthesis over the conservative baseline, with larger values indicating a greater reduction in disclosure risk. In addition to the risk reduction across the entire confidential dataset, we also compute it with reference to unique individuals by calculating a bootstrap averaged $\overline{CMAP_{s_{unique}}^{\epsilon}}$. We determine an individual $j$ is unique if $\sum_{i=1}^{n}\left[\boldsymbol{K_{oj} = K_{oi}}\right] = 1$. In the presence of an overfit synthesizer, individuals with outlying characteristics are at highest risk for sensitive attribute disclosure. 

\begin{figure}[H]
    \centering
    \includegraphics[width = .49\textwidth, keepaspectratio]{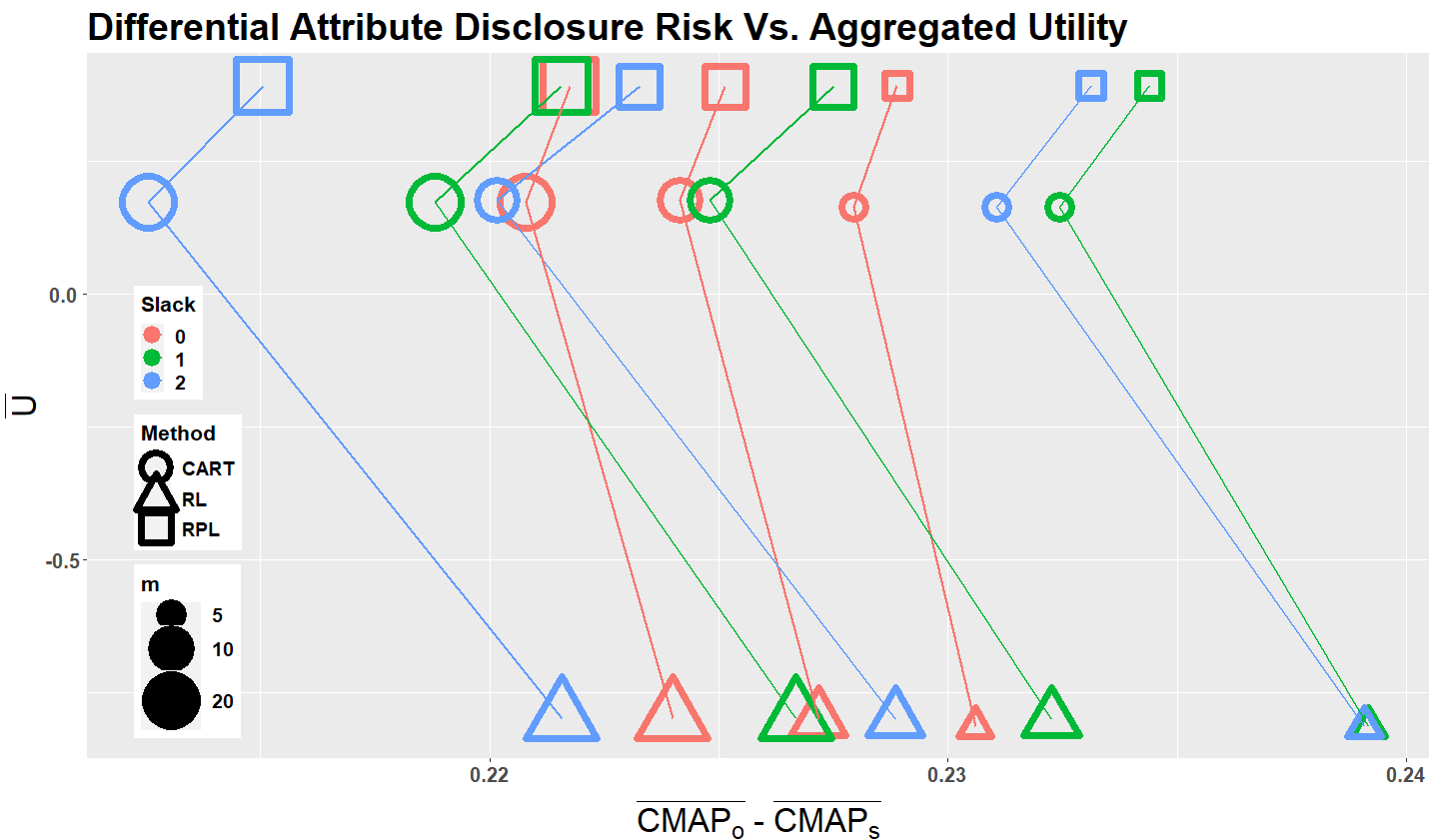}
    \includegraphics[width = .49\textwidth, keepaspectratio]{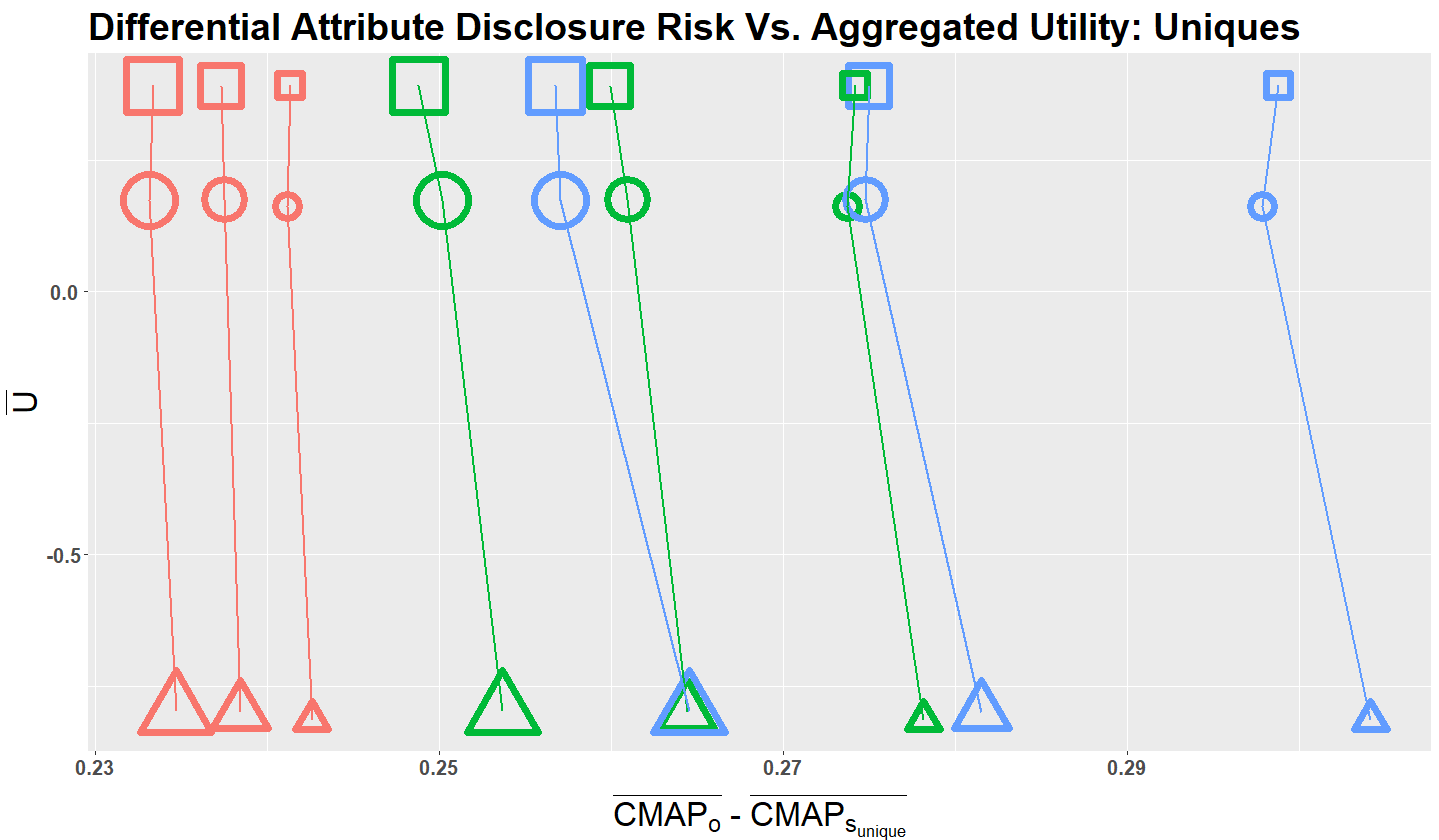}
    \caption{The utility-risk trade-off for competing synthesizers, varying data dissemination parameters: lines are drawn to facilitate comparison between methods and across pairs of $\epsilon$ and $m$. In the left panel, we observe that RPL synthesis simultaneously provides greater differential attribute disclosure risk reduction and higher utility than CART for each $\epsilon$ and $m$. For uniques, differential risk reductions are approximately equivalent, demonstrating the superior utility of RPL synthetic datasets. As for data dissemination, releasing more datasets results in higher disclosure risk, while utility remains approximately the same. }
    \label{fig7}
\end{figure}

The differential risk reduction for the confidential dataset and uniques are plotted against the mean aggregated utility $\bar{U}$ in Figure~\ref{fig7}. For each sample of $m$ datasets, we compute $\bar{U} = \frac{1}{m}\sum_{i = 1}^{m} U_{i}$ and average this across bootstrap samples. The aggregated utility will only vary as a function of $m$. In Figure~\ref{fig7}, we have further fixed $\lvert \boldsymbol{K_{o}}\rvert = 7$. This choice is motivated by Figures~\ref{Risk}-\ref{Risk2}, which suggest high baseline attribute disclosure risks and that the risk reduction is greatest when the intruder is assumed to have this level of prior knowledge on the confidential set.

Because $\overline{CMAP_{o}^{\epsilon}}$ is equivalent across synthesis methods for each $\epsilon$, the risk reductions can be directly compared. When evaluating disclosure risk reduction for the entire confidential dataset (left panel of Figure~\ref{fig7}), it is clear that RPL synthesis provides a greater reduction than CART over the baseline for each combination of $\epsilon$ and $m$. For uniques, the disclosure risks are approximately equivalent across methods, but the risk reduction is greater (relative to the entire population), suggesting increased protection for these individuals.

Furthermore, RPL synthetic data has the highest mean aggregated utility --- the square icons are highest along the y-axis for each combination of $\epsilon$ and $m$. Therefore, under the assumed amount of intruder knowledge and proposed dissemination procedure, the RPL is uniformly preferable over CART in terms of the utility-risk trade-off.

As for the process of releasing synthetic datasets publicly, the results in Figure~\ref{fig7} are consistent with \cite{reiter2009estimating}. Applicable to each method, the size of the icons increase as the differential risk reduction decreases. This demonstrates increased disclosure risk as more synthetic datasets are released. Coupled with negligible gains in aggregated utility --- the icons basically remain on the same y-axis line --- this suggests that the data disseminating agency should closely consider any disclosure risk criteria prior to releasing large numbers of synthetic datasets.

\section{Conclusion}\label{conclude}
We have developed a Bayesian semiparametric modeling framework for mixed categorical, binary, count, and continuous variables. The joint Bayesian model and  targeted univariate synthesis of candidate response variables in focal regression models   are particularly useful for generating reliable and privacy-preserving synthetic datasets based on sensitive micro data. 

Micro data collected on individuals are essential for health and health disparities research and practice, yet cannot be released due to privacy concerns. This undermines scientific reproducibility and inhibits further study of critical questions about public health and child development. The proposed approach remedies theses issues by generating synthetic datasets that empirically capture marginal, bivariate, and (nonlinear) regression associations among variables, while limiting privacy concerns from re-indentification and attribute disclosure risks. Unlike existing Bayesian methods for data synthesis, the proposed \emph{extended rank-probit likelihood} incorporates unordered categorical variables---alongside binary, count, and continuous variables---and can synthesize variables without needing to pre-select a variable ordering. Since regression analysis is a common tool in epidemiological studies, we introduce a modified data synthesis strategy to target and preserve key conditional relationships, including both nonlinearities and interactions.

Comparing the proposed methods with Bayesian and non-Bayesian competitors, we repeatedly generate synthetic versions of a confidential dataset containing dozens of demographic, socioeconomic, environmental exposure, and educational outcome measurements on nearly 20,000 North Carolina children. We highlight the stable inference for regression models of interest under our framework by pooling estimates from each synthetic dataset and constructing confidence intervals. When compared to the other methods, our synthesis model provided more consistent point estimation and uncertainty quantificationg for regression coefficients.

Using a new attribute disclosure risk metric, we then investigate the utility-risk trade-off of fully synthetic data as a function of several \textit{post-hoc} data-dissemination decisions. We demonstrate that all methods produce synthetic data with similarly low attribute disclosure in comparison to a conservative baseline. On average, the synthetic data produced under our method had the highest utility, regardless of the number of datasets released. Assuming conservative levels of intruder prior knowledge, our method was also shown to provide a more substantial disclosure risk reduction over baseline than CART.

 Compared to the confidential version, our framework produces synthetic versions that matches marginal distributions, even for highly non-Gaussian variables; reproduces cross-tabulations between categorical variables; preserves posterior inference for linear regressions with shrinkage priors and interactions; and limits both re-indentification and attribute disclosure risks. The proposed approach demonstrates substantial improvements over existing Bayesian methods, especially for modeling of unordered categorical variables and ordinal variables with few levels, and competitive performance with a popular nonparametric synthesizer in both utility and privacy preservation. The role of unordered categorical variables---in conjunction with binary, count, and continuous variables---cannot be understated: race or ethnicity is one such variable, and is a critical factor in health and health disparities research and practice.

Although we have deployed the extended rank-probit likelihood factor model for data synthesis, this Bayesian modeling framework may also be useful for inference on mixed data. For instance, the copula correlation matrices $\boldsymbol{C}$ offer a new way to quantify associations among categorical, binary, count, and continuous variables. 

In addition, the extended rank-probit likelihood provides a natural extension of multinomial choice models. Specifically, the diagonal orthant probit multinomial model of \cite{johndrow2013diagonal} relies on a diagonal covariance for model fitting, translating to an assumption of independence of irrelevant alternatives. The extended rank-probit likelihood, coupled with the factor model \eqref{eq22}, offers an opportunity to relax this assumption. 

Beyond these inferential extensions, future work will investigate the possibility of incorporating nonlinearity into the Gaussian copula model. For instance, replacing \eqref{eq22} with a nonlinear factor model may offer similar advantages as the nonlinear regression in Section~\ref{nonlinear}, yet without the need to single out the response variables of interest. However, additional modeling complexity will introduce new computational challenges, and will require special care to avoid overfitting---which can lead to greater attribute disclosure risks.

\section*{Acknowledgements}
The authors would like to thank the reviewers for their constructive comments that have greatly improved the paper. In addition, the authors thank Marie Lynn Miranda and Katherine B. Ensor for their valuable insights and feedback. Research reported in this publication was supported by the National Institute of Environmental Health Sciences of the National Institutes of Health under award number R01ES028819 and the Army Research Office (Kowal) under award number W911NF-20-1-0184. The content, views, and conclusions contained in this document are those of the authors and should not be interpreted as representing the official policies, either expressed or implied, of the National Institutes of Health, the North Carolina Department of Health and Human Services, Division of Public Health, the Army Research Office, or the U.S. Government. The U.S. Government is authorized to reproduce and distribute reprints for Government purposes notwithstanding any copyright notation herein.

\begin{supplement}
\stitle{Detailed dataset description, additional model details, and more simulation results}
\sdescription{A. Detailed Dataset Description  B. The RPL and continuous data  C. Specification of hyperparameters for the priors in section \ref{sec4} D. Additional Simulation E. Utility-Risk plots for different levels of intruder prior knowledge, akin to Figure~\ref{fig7}}
\end{supplement}
\begin{supplement}
\stitle{R code}
\sdescription{Example R code for synthesis of a simulated data set containing categorical, binary, continuous, and count variables. Highlighting our strategy, the count variable is simulated as a non-linear function of the other variables, and is isolated for targeted synthesis in this case.  The MCMC algorithm, copula synthesizer, helper functions, and nonlinear synthesizer (with basic diagnostic tools) are included in a zip file.}
\end{supplement}
\bibliographystyle{imsart-nameyear} 
\nocite{*}
\bibliography{Bib.bib}       

\begin{thebibliography}{42}

\bibitem[\protect\citeauthoryear{Albert and Chib}{1993}]{albert1993bayesian}
\begin{barticle}[author]
\bauthor{\bsnm{Albert},~\bfnm{James~H}\binits{J.~H.}} \AND
  \bauthor{\bsnm{Chib},~\bfnm{Siddhartha}\binits{S.}}
(\byear{1993}).
\btitle{Bayesian analysis of binary and polychotomous response data}.
\bjournal{Journal of the American statistical Association}
\bvolume{88}
\bpages{669--679}.
\end{barticle}
\endbibitem

\bibitem[\protect\citeauthoryear{Bernardo et~al.}{2003}]{bernardo2003bayesian}
\begin{barticle}[author]
\bauthor{\bsnm{Bernardo},~\bfnm{JM}\binits{J.}},
  \bauthor{\bsnm{Bayarri},~\bfnm{MJ}\binits{M.}},
  \bauthor{\bsnm{Berger},~\bfnm{JO}\binits{J.}},
  \bauthor{\bsnm{Dawid},~\bfnm{AP}\binits{A.}},
  \bauthor{\bsnm{Heckerman},~\bfnm{D}\binits{D.}},
  \bauthor{\bsnm{Smith},~\bfnm{A}\binits{A.}} \AND
  \bauthor{\bsnm{West},~\bfnm{M}\binits{M.}}
(\byear{2003}).
\btitle{Bayesian factor regression models in the “large p, small n”
  paradigm}.
\bjournal{Bayesian statistics}
\bvolume{7}
\bpages{733--742}.
\end{barticle}
\endbibitem

\bibitem[\protect\citeauthoryear{Bhattacharya and
  Dunson}{2011}]{bhattacharya2011sparse}
\begin{barticle}[author]
\bauthor{\bsnm{Bhattacharya},~\bfnm{Anirban}\binits{A.}} \AND
  \bauthor{\bsnm{Dunson},~\bfnm{David~B}\binits{D.~B.}}
(\byear{2011}).
\btitle{Sparse Bayesian infinite factor models}.
\bjournal{Biometrika}
\bpages{291--306}.
\end{barticle}
\endbibitem

\bibitem[\protect\citeauthoryear{Caiola and Reiter}{2010}]{caiola2010random}
\begin{barticle}[author]
\bauthor{\bsnm{Caiola},~\bfnm{Gregory}\binits{G.}} \AND
  \bauthor{\bsnm{Reiter},~\bfnm{Jerome~P}\binits{J.~P.}}
(\byear{2010}).
\btitle{Random forests for generating partially synthetic, categorical data.}
\bjournal{Trans. Data Priv.}
\bvolume{3}
\bpages{27--42}.
\end{barticle}
\endbibitem

\bibitem[\protect\citeauthoryear{Carvalho, Polson and
  Scott}{2010}]{carvalho2010horseshoe}
\begin{barticle}[author]
\bauthor{\bsnm{Carvalho},~\bfnm{Carlos~M}\binits{C.~M.}},
  \bauthor{\bsnm{Polson},~\bfnm{Nicholas~G}\binits{N.~G.}} \AND
  \bauthor{\bsnm{Scott},~\bfnm{James~G}\binits{J.~G.}}
(\byear{2010}).
\btitle{The horseshoe estimator for sparse signals}.
\bjournal{Biometrika}
\bvolume{97}
\bpages{465--480}.
\end{barticle}
\endbibitem

\bibitem[\protect\citeauthoryear{Chipman, George and
  McCulloch}{2010}]{chipman2010bart}
\begin{barticle}[author]
\bauthor{\bsnm{Chipman},~\bfnm{Hugh~A}\binits{H.~A.}},
  \bauthor{\bsnm{George},~\bfnm{Edward~I}\binits{E.~I.}} \AND
  \bauthor{\bsnm{McCulloch},~\bfnm{Robert~E}\binits{R.~E.}}
(\byear{2010}).
\btitle{BART: Bayesian additive regression trees}.
\bjournal{The Annals of Applied Statistics}
\bvolume{4}
\bpages{266--298}.
\end{barticle}
\endbibitem

\bibitem[\protect\citeauthoryear{Drechsler}{2018}]{drechsler2018some}
\begin{binproceedings}[author]
\bauthor{\bsnm{Drechsler},~\bfnm{J{\"o}rg}\binits{J.}}
(\byear{2018}).
\btitle{Some clarifications regarding fully synthetic data}.
In \bbooktitle{International Conference on Privacy in Statistical Databases}
\bpages{109--121}.
\bpublisher{Springer}.
\end{binproceedings}
\endbibitem

\bibitem[\protect\citeauthoryear{Drechsler and
  Reiter}{2011}]{drechsler2011empirical}
\begin{barticle}[author]
\bauthor{\bsnm{Drechsler},~\bfnm{J{\"o}rg}\binits{J.}} \AND
  \bauthor{\bsnm{Reiter},~\bfnm{Jerome~P}\binits{J.~P.}}
(\byear{2011}).
\btitle{An empirical evaluation of easily implemented, nonparametric methods
  for generating synthetic datasets}.
\bjournal{Computational Statistics \& Data Analysis}
\bvolume{55}
\bpages{3232--3243}.
\end{barticle}
\endbibitem

\bibitem[\protect\citeauthoryear{Duncan, Keller-McNulty and
  Stokes}{2001}]{Duncan2001}
\begin{barticle}[author]
\bauthor{\bsnm{Duncan},~\bfnm{G.~T.}\binits{G.~T.}},
  \bauthor{\bsnm{Keller-McNulty},~\bfnm{S.~A}\binits{S.~A.}} \AND
  \bauthor{\bsnm{Stokes},~\bfnm{S.~L}\binits{S.~L.}}
(\byear{2001}).
\btitle{Disclosure Risk Vs Data Utility: The R-U Confidentiality Map}.
\end{barticle}
\endbibitem

\bibitem[\protect\citeauthoryear{Dunson and
  Xing}{2009}]{dunson2009nonparametric}
\begin{barticle}[author]
\bauthor{\bsnm{Dunson},~\bfnm{David~B}\binits{D.~B.}} \AND
  \bauthor{\bsnm{Xing},~\bfnm{Chuanhua}\binits{C.}}
(\byear{2009}).
\btitle{Nonparametric Bayes modeling of multivariate categorical data}.
\bjournal{Journal of the American Statistical Association}
\bvolume{104}
\bpages{1042--1051}.
\end{barticle}
\endbibitem

\bibitem[\protect\citeauthoryear{Elliot}{2015}]{elliot2015final}
\begin{barticle}[author]
\bauthor{\bsnm{Elliot},~\bfnm{Mark}\binits{M.}}
(\byear{2015}).
\btitle{Final report on the disclosure risk associated with the synthetic data
  produced by the sylls team}.
\bjournal{Report 2015}
\bvolume{2}.
\end{barticle}
\endbibitem

\bibitem[\protect\citeauthoryear{Feldman and Kowal}{2022}]{supp}
\begin{barticle}[author]
\bauthor{\bsnm{Feldman},~\bfnm{Joseph}\binits{J.}} \AND
  \bauthor{\bsnm{Kowal},~\bfnm{Daniel}\binits{D.}}
(\byear{2022}).
\btitle{Supplement to “Bayesian Data Synthesis and the Utility-Risk Trade-Off
  for Mixed Epidemiological Data”}.
\end{barticle}
\endbibitem

\bibitem[\protect\citeauthoryear{Ferrari and
  Dunson}{2020}]{ferrari2020bayesian}
\begin{barticle}[author]
\bauthor{\bsnm{Ferrari},~\bfnm{Federico}\binits{F.}} \AND
  \bauthor{\bsnm{Dunson},~\bfnm{David~B}\binits{D.~B.}}
(\byear{2020}).
\btitle{Bayesian factor analysis for inference on interactions}.
\bjournal{Journal of the American Statistical Association}
\bpages{1--12}.
\end{barticle}
\endbibitem

\bibitem[\protect\citeauthoryear{Hoff}{2007}]{hoff2007extending}
\begin{barticle}[author]
\bauthor{\bsnm{Hoff},~\bfnm{Peter~D}\binits{P.~D.}}
(\byear{2007}).
\btitle{Extending the rank likelihood for semiparametric copula estimation}.
\bjournal{The Annals of Applied Statistics}
\bvolume{1}
\bpages{265--283}.
\end{barticle}
\endbibitem

\bibitem[\protect\citeauthoryear{Hu}{2019}]{Hu2019}
\begin{barticle}[author]
\bauthor{\bsnm{Hu},~\bfnm{Jingchen}\binits{J.}}
(\byear{2019}).
\btitle{Bayesian Estimation of Attribute and Identification Disclosure Risks in
  Synthetic Data}.
\bpages{61--89}.
\end{barticle}
\endbibitem

\bibitem[\protect\citeauthoryear{Hu, Reiter and Wang}{2014}]{hu2014disclosure}
\begin{binproceedings}[author]
\bauthor{\bsnm{Hu},~\bfnm{Jingchen}\binits{J.}},
  \bauthor{\bsnm{Reiter},~\bfnm{Jerome~P}\binits{J.~P.}} \AND
  \bauthor{\bsnm{Wang},~\bfnm{Quanli}\binits{Q.}}
(\byear{2014}).
\btitle{Disclosure risk evaluation for fully synthetic categorical data}.
In \bbooktitle{International conference on privacy in statistical databases}
\bpages{185--199}.
\bpublisher{Springer}.
\end{binproceedings}
\endbibitem

\bibitem[\protect\citeauthoryear{Johndrow, Dunson and
  Lum}{2013}]{johndrow2013diagonal}
\begin{binproceedings}[author]
\bauthor{\bsnm{Johndrow},~\bfnm{James}\binits{J.}},
  \bauthor{\bsnm{Dunson},~\bfnm{David}\binits{D.}} \AND
  \bauthor{\bsnm{Lum},~\bfnm{Kristian}\binits{K.}}
(\byear{2013}).
\btitle{Diagonal orthant multinomial probit models}.
In \bbooktitle{Artificial Intelligence and Statistics}
\bpages{29--38}.
\end{binproceedings}
\endbibitem

\bibitem[\protect\citeauthoryear{Kinney et~al.}{2011}]{kinney2011towards}
\begin{barticle}[author]
\bauthor{\bsnm{Kinney},~\bfnm{Satkartar~K}\binits{S.~K.}},
  \bauthor{\bsnm{Reiter},~\bfnm{Jerome~P}\binits{J.~P.}},
  \bauthor{\bsnm{Reznek},~\bfnm{Arnold~P}\binits{A.~P.}},
  \bauthor{\bsnm{Miranda},~\bfnm{Javier}\binits{J.}},
  \bauthor{\bsnm{Jarmin},~\bfnm{Ron~S}\binits{R.~S.}} \AND
  \bauthor{\bsnm{Abowd},~\bfnm{John~M}\binits{J.~M.}}
(\byear{2011}).
\btitle{Towards unrestricted public use business microdata: The synthetic
  longitudinal business database}.
\bjournal{International Statistical Review}
\bvolume{79}
\bpages{362--384}.
\end{barticle}
\endbibitem

\bibitem[\protect\citeauthoryear{Kowal}{2021}]{kowal2021fast}
\begin{barticle}[author]
\bauthor{\bsnm{Kowal},~\bfnm{Daniel~R}\binits{D.~R.}}
(\byear{2021}).
\btitle{Fast, Optimal, and Targeted Predictions Using Parameterized Decision
  Analysis}.
\bjournal{Journal of the American Statistical Association}
\bpages{1--12}.
\end{barticle}
\endbibitem

\bibitem[\protect\citeauthoryear{Kowal and
  Canale}{2020}]{kowal2020simultaneous}
\begin{barticle}[author]
\bauthor{\bsnm{Kowal},~\bfnm{Daniel~R}\binits{D.~R.}} \AND
  \bauthor{\bsnm{Canale},~\bfnm{Antonio}\binits{A.}}
(\byear{2020}).
\btitle{Simultaneous transformation and rounding (STAR) models for
  integer-valued data}.
\bjournal{Electronic Journal of Statistics}
\bvolume{14}
\bpages{1744--1772}.
\end{barticle}
\endbibitem

\bibitem[\protect\citeauthoryear{Kowal et~al.}{2020}]{BVS}
\begin{barticle}[author]
\bauthor{\bsnm{Kowal},~\bfnm{Daniel~R.}\binits{D.~R.}},
  \bauthor{\bsnm{Bravo},~\bfnm{Mercedes}\binits{M.}},
  \bauthor{\bsnm{Leong},~\bfnm{Henry}\binits{H.}},
  \bauthor{\bsnm{Griffin},~\bfnm{Robert~J.}\binits{R.~J.}},
  \bauthor{\bsnm{Ensor},~\bfnm{Katherine~B.}\binits{K.~B.}} \AND
  \bauthor{\bsnm{Miranda},~\bfnm{Marie~Lynn}\binits{M.~L.}}
(\byear{2020}).
\btitle{{Bayesian Variable Selection for Understanding Mixtures in
  Environmental Exposures}}.
\bjournal{Statistics in Medicine}.
\end{barticle}
\endbibitem

\bibitem[\protect\citeauthoryear{Little}{1993}]{little1993statistical}
\begin{barticle}[author]
\bauthor{\bsnm{Little},~\bfnm{Roderick~JA}\binits{R.~J.}}
(\byear{1993}).
\btitle{Statistical analysis of masked data}.
\bjournal{Journal of Official statistics}
\bvolume{9}
\bpages{407}.
\end{barticle}
\endbibitem

\bibitem[\protect\citeauthoryear{Miranda
  et~al.}{2007}]{miranda2007relationship}
\begin{barticle}[author]
\bauthor{\bsnm{Miranda},~\bfnm{Marie~Lynn}\binits{M.~L.}},
  \bauthor{\bsnm{Kim},~\bfnm{Dohyeong}\binits{D.}},
  \bauthor{\bsnm{Galeano},~\bfnm{M~Alicia~Overstreet}\binits{M.~A.~O.}},
  \bauthor{\bsnm{Paul},~\bfnm{Christopher~J}\binits{C.~J.}},
  \bauthor{\bsnm{Hull},~\bfnm{Andrew~P}\binits{A.~P.}} \AND
  \bauthor{\bsnm{Morgan},~\bfnm{S~Philip}\binits{S.~P.}}
(\byear{2007}).
\btitle{The relationship between early childhood blood lead levels and
  performance on end-of-grade tests}.
\bjournal{Environmental Health Perspectives}
\bvolume{115}
\bpages{1242--1247}.
\end{barticle}
\endbibitem

\bibitem[\protect\citeauthoryear{Murray and Reiter}{2016}]{murray2016multiple}
\begin{barticle}[author]
\bauthor{\bsnm{Murray},~\bfnm{Jared~S}\binits{J.~S.}} \AND
  \bauthor{\bsnm{Reiter},~\bfnm{Jerome~P}\binits{J.~P.}}
(\byear{2016}).
\btitle{Multiple imputation of missing categorical and continuous values via
  Bayesian mixture models with local dependence}.
\bjournal{Journal of the American Statistical Association}
\bvolume{111}
\bpages{1466--1479}.
\end{barticle}
\endbibitem

\bibitem[\protect\citeauthoryear{Murray et~al.}{2013}]{murray2013bayesian}
\begin{barticle}[author]
\bauthor{\bsnm{Murray},~\bfnm{Jared~S}\binits{J.~S.}},
  \bauthor{\bsnm{Dunson},~\bfnm{David~B}\binits{D.~B.}},
  \bauthor{\bsnm{Carin},~\bfnm{Lawrence}\binits{L.}} \AND
  \bauthor{\bsnm{Lucas},~\bfnm{Joseph~E}\binits{J.~E.}}
(\byear{2013}).
\btitle{Bayesian Gaussian copula factor models for mixed data}.
\bjournal{Journal of the American Statistical Association}
\bvolume{108}
\bpages{656--665}.
\end{barticle}
\endbibitem

\bibitem[\protect\citeauthoryear{Nowok}{2015}]{nowok2015utility}
\begin{barticle}[author]
\bauthor{\bsnm{Nowok},~\bfnm{Beata}\binits{B.}}
(\byear{2015}).
\btitle{Utility of synthetic microdata generated using tree-based methods}.
\bjournal{UNECE Statistical Data Confidentiality Work Session}.
\end{barticle}
\endbibitem

\bibitem[\protect\citeauthoryear{Nowok et~al.}{2016}]{nowok2016synthpop}
\begin{barticle}[author]
\bauthor{\bsnm{Nowok},~\bfnm{B}\binits{B.}},
  \bauthor{\bsnm{Raab},~\bfnm{GM}\binits{G.}},
  \bauthor{\bsnm{Snoke},~\bfnm{J}\binits{J.}} \AND
  \bauthor{\bsnm{Dibben},~\bfnm{C}\binits{C.}}
(\byear{2016}).
\btitle{synthpop: Generating Synthetic Versions of Sensitive Microdata for
  Statistical Disclosure Control}.
\bjournal{R package version}
\bpages{1--3}.
\end{barticle}
\endbibitem

\bibitem[\protect\citeauthoryear{Quick et~al.}{2015}]{quick2015bayesian}
\begin{barticle}[author]
\bauthor{\bsnm{Quick},~\bfnm{Harrison}\binits{H.}},
  \bauthor{\bsnm{Holan},~\bfnm{Scott~H}\binits{S.~H.}},
  \bauthor{\bsnm{Wikle},~\bfnm{Christopher~K}\binits{C.~K.}} \AND
  \bauthor{\bsnm{Reiter},~\bfnm{Jerome~P}\binits{J.~P.}}
(\byear{2015}).
\btitle{Bayesian marked point process modeling for generating fully synthetic
  public use data with point-referenced geography}.
\bjournal{Spatial Statistics}
\bvolume{14}
\bpages{439--451}.
\end{barticle}
\endbibitem

\bibitem[\protect\citeauthoryear{Quinn}{2004}]{quinn2004bayesian}
\begin{barticle}[author]
\bauthor{\bsnm{Quinn},~\bfnm{Kevin~M}\binits{K.~M.}}
(\byear{2004}).
\btitle{Bayesian factor analysis for mixed ordinal and continuous responses}.
\bjournal{Political Analysis}
\bvolume{12}
\bpages{338--353}.
\end{barticle}
\endbibitem

\bibitem[\protect\citeauthoryear{Raab, Nowok and
  Dibben}{2016}]{raab2016practical}
\begin{barticle}[author]
\bauthor{\bsnm{Raab},~\bfnm{Gillian~M}\binits{G.~M.}},
  \bauthor{\bsnm{Nowok},~\bfnm{Beata}\binits{B.}} \AND
  \bauthor{\bsnm{Dibben},~\bfnm{Chris}\binits{C.}}
(\byear{2016}).
\btitle{Practical data synthesis for large samples}.
\bjournal{Journal of Privacy and Confidentiality}
\bvolume{7}
\bpages{67--97}.
\end{barticle}
\endbibitem

\bibitem[\protect\citeauthoryear{Raghunathan, Reiter and
  Rubin}{2003}]{raghunathan2003multiple}
\begin{barticle}[author]
\bauthor{\bsnm{Raghunathan},~\bfnm{Trivellore~E}\binits{T.~E.}},
  \bauthor{\bsnm{Reiter},~\bfnm{Jerome~P}\binits{J.~P.}} \AND
  \bauthor{\bsnm{Rubin},~\bfnm{Donald~B}\binits{D.~B.}}
(\byear{2003}).
\btitle{Multiple imputation for statistical disclosure limitation}.
\bjournal{Journal of official statistics}
\bvolume{19}
\bpages{1}.
\end{barticle}
\endbibitem

\bibitem[\protect\citeauthoryear{Reiter}{2003}]{Reiter2003}
\begin{barticle}[author]
\bauthor{\bsnm{Reiter},~\bfnm{J.~P.}\binits{J.~P.}}
(\byear{2003}).
\btitle{Inference for partially synthetic, public use microdata sets}.
\bjournal{Survey Methodology}
\bvolume{29}
\bpages{181-188}.
\end{barticle}
\endbibitem

\bibitem[\protect\citeauthoryear{Reiter}{2005a}]{reiter2005releasing}
\begin{barticle}[author]
\bauthor{\bsnm{Reiter},~\bfnm{Jerome~P}\binits{J.~P.}}
(\byear{2005}a).
\btitle{Releasing multiply imputed, synthetic public use microdata: an
  illustration and empirical study}.
\bjournal{Journal of the Royal Statistical Society: Series A (Statistics in
  Society)}
\bvolume{168}
\bpages{185--205}.
\end{barticle}
\endbibitem

\bibitem[\protect\citeauthoryear{Reiter}{2005b}]{reiter2005using}
\begin{barticle}[author]
\bauthor{\bsnm{Reiter},~\bfnm{Jerome~P}\binits{J.~P.}}
(\byear{2005}b).
\btitle{Using CART to generate partially synthetic public use microdata}.
\bjournal{Journal of Official Statistics}
\bvolume{21}
\bpages{441}.
\end{barticle}
\endbibitem

\bibitem[\protect\citeauthoryear{Reiter and Mitra}{2009}]{reiter2009estimating}
\begin{barticle}[author]
\bauthor{\bsnm{Reiter},~\bfnm{Jerome~P}\binits{J.~P.}} \AND
  \bauthor{\bsnm{Mitra},~\bfnm{Robin}\binits{R.}}
(\byear{2009}).
\btitle{Estimating risks of identification disclosure in partially synthetic
  data}.
\bjournal{Journal of Privacy and Confidentiality}
\bvolume{1}.
\end{barticle}
\endbibitem

\bibitem[\protect\citeauthoryear{Reiter, Wang and
  Zhang}{2014}]{reiter2014bayesian}
\begin{barticle}[author]
\bauthor{\bsnm{Reiter},~\bfnm{Jerome~P}\binits{J.~P.}},
  \bauthor{\bsnm{Wang},~\bfnm{Quanli}\binits{Q.}} \AND
  \bauthor{\bsnm{Zhang},~\bfnm{Biyuan}\binits{B.}}
(\byear{2014}).
\btitle{Bayesian estimation of disclosure risks for multiply imputed, synthetic
  data}.
\bjournal{Journal of Privacy and Confidentiality}
\bvolume{6}.
\end{barticle}
\endbibitem

\bibitem[\protect\citeauthoryear{Rubin}{1993}]{rubin1993statistical}
\begin{barticle}[author]
\bauthor{\bsnm{Rubin},~\bfnm{Donald~B}\binits{D.~B.}}
(\byear{1993}).
\btitle{Statistical disclosure limitation}.
\bjournal{Journal of official Statistics}
\bvolume{9}
\bpages{461--468}.
\end{barticle}
\endbibitem

\bibitem[\protect\citeauthoryear{Sklar}{1959}]{sklar1959fonctions}
\begin{barticle}[author]
\bauthor{\bsnm{Sklar},~\bfnm{M}\binits{M.}}
(\byear{1959}).
\btitle{Fonctions de repartition an dimensions et leurs marges}.
\bjournal{Publ. inst. statist. univ. Paris}
\bvolume{8}
\bpages{229--231}.
\end{barticle}
\endbibitem

\bibitem[\protect\citeauthoryear{Snoke et~al.}{2018}]{snoke2018general}
\begin{barticle}[author]
\bauthor{\bsnm{Snoke},~\bfnm{Joshua}\binits{J.}},
  \bauthor{\bsnm{Raab},~\bfnm{Gillian~M}\binits{G.~M.}},
  \bauthor{\bsnm{Nowok},~\bfnm{Beata}\binits{B.}},
  \bauthor{\bsnm{Dibben},~\bfnm{Chris}\binits{C.}} \AND
  \bauthor{\bsnm{Slavkovic},~\bfnm{Aleksandra}\binits{A.}}
(\byear{2018}).
\btitle{General and specific utility measures for synthetic data}.
\bjournal{Journal of the Royal Statistical Society. Series A: Statistics in
  Society}
\bvolume{181}
\bpages{663--688}.
\end{barticle}
\endbibitem

\bibitem[\protect\citeauthoryear{Taub}{2021}]{taub2021synthetic}
\begin{barticle}[author]
\bauthor{\bsnm{Taub},~\bfnm{Jennifer}\binits{J.}}
(\byear{2021}).
\btitle{Synthetic Data: An Exploration of Data Utility and Disclosure Risk}.
\bpages{138--160}.
\end{barticle}
\endbibitem

\bibitem[\protect\citeauthoryear{Taub et~al.}{2018}]{taub2018differential}
\begin{binproceedings}[author]
\bauthor{\bsnm{Taub},~\bfnm{Jennifer}\binits{J.}},
  \bauthor{\bsnm{Elliot},~\bfnm{Mark}\binits{M.}},
  \bauthor{\bsnm{Pampaka},~\bfnm{Maria}\binits{M.}} \AND
  \bauthor{\bsnm{Smith},~\bfnm{Duncan}\binits{D.}}
(\byear{2018}).
\btitle{Differential correct attribution probability for synthetic data: An
  Exploration}.
In \bbooktitle{International Conference on Privacy in Statistical Databases}
\bpages{122--137}.
\bpublisher{Springer}.
\end{binproceedings}
\endbibitem

\bibitem[\protect\citeauthoryear{Woo et~al.}{2009}]{woo2009global}
\begin{barticle}[author]
\bauthor{\bsnm{Woo},~\bfnm{Mi-Ja}\binits{M.-J.}},
  \bauthor{\bsnm{Reiter},~\bfnm{Jerome~P}\binits{J.~P.}},
  \bauthor{\bsnm{Oganian},~\bfnm{Anna}\binits{A.}} \AND
  \bauthor{\bsnm{Karr},~\bfnm{Alan~F}\binits{A.~F.}}
(\byear{2009}).
\btitle{Global measures of data utility for microdata masked for disclosure
  limitation}.
\bjournal{Journal of Privacy and Confidentiality}
\bvolume{1}.
\end{barticle}
\endbibitem

\end{thebibliography}

\end{document}